\documentclass[aps,prb,floatfix,twocolumn,amsmath,amssymb,superscriptaddress]{revtex4-1}
\usepackage{graphicx}
\usepackage{hyperref, color, cancel, comment,bbold,bm}
\newcommand{\be}{\begin{equation}}
\newcommand{\ben}{\begin{equation*}}
\newcommand{\ee}{\end{equation}}
\newcommand{\een}{\end{equation*}}
\newcommand{\bs}{\begin{split}}
\newcommand{\es}{\end{split}}
\newcommand{\bmx}{\begin{array}}
\newcommand{\emx}{\end{array}}
\newcommand{\bea}{\begin{eqnarray}}
\newcommand{\bean}{\begin{eqnarray*}}
\newcommand{\eea}{\end{eqnarray}}
\newcommand{\eean}{\end{eqnarray*}}

\newcommand{\dn}{^{\vphantom{\dagger}}}

\newcommand{\bb}[1]{\mathbb{#1}}

\newcommand{\so}{\qquad\rightarrow\qquad}

\newcommand{\eps}{\epsilon}

\newcommand{\sgn}[1]{{\rm sign}{#1}}
\newcommand{\pref}[1]{(\ref{#1})}

\newcommand{\abs}[1]{\left\vert #1 \right\vert}

\newcommand{\com}[2]{\left[#1,#2\right]}

\newcommand{\mat}[1]{\left(\bmx{cc}#1\emx\right)}

\usepackage{chngcntr}
\counterwithin{equation}{section}
\renewcommand\theequation{\arabic{section}.\arabic{equation}}
\renewcommand{\inf}{\infty}
\begin{document}
\title{Luttinger liquid in contact with a Kramers pair of Majorana bound states}

\author{Dmitry I. Pikulin,$^{1,2}$  Yashar Komijani$^{1,2,3}$ and Ian Affleck}
\affiliation{Department of Physics and Astronomy and 
$^2$Quantum Matter Institute, University of British Columbia, Vancouver, BC, Canada V6T 1Z1\\
$^3$Center for Materials Theory, Rutgers University,
Piscataway, New Jersey, 08854, USA}
\date{\today}

\begin{abstract}
We discuss the signatures of a Kramers pair of Majorana modes formed in a Josephson junction on top of a quantum spin Hall system. 
We show that, while ignoring interactions on the quantum spin Hall edge allows arbitrary Andreev process in the system, moderate repulsive interactions stabilize Andreev transmission --  the hole goes into the \textit{opposite} lead from where the electron has arrived. We analyze the renormalization group equations and deduce the existence of a non-trivial critical point for sufficiently strong interactions. 
\end{abstract}
\maketitle

\section{Introduction}

Majorana quasi-particles have been in the focus of condensed matter research in recent years.\cite{Ali12,Bee13} The promise of topologically protected memory, quantum computation, and non-abelian statistics \cite{Nay08} has drawn the attention of both theorists and experimentalists.

The first signatures of  Majoranas -- zero-bias differential conductance peaks -- have been reported in semiconductor nanowires\cite{Lut10, Ore10, Mou12, Das12, Den12} and chains of magnetic atoms in contact with bulk superconductors.\cite{Cho11, Nad13, Nad14} Additional theoretical proposals include proximitized 2-dimensional\cite{LFu09, Bee12, SMi13} and 3-dimensional topological insulators.\cite{LFu08}  Much more investigation is necessary since the zero-bias peak can have multiple other origins, including 
disorder.\cite{Ios12, Bag12, Pik12}

Recently the possibility of obtaining multiple Majoranas located in the same physical position but not hybridizing with each other due to additional symmetry has been put forward. \cite{10foldway, Teo10, Tew12, Zha13a, Zha13b, Kes13, Zha14, Gai14, Hai14, Kli14, Kli14a, Sch15} Besides widening our understanding of the topological phases, observation of such system will strengthen the claim of the Majorana bound states\cite{Fid10, Fid11, Tur11, Chi15, Pik15} even before braiding -- an unparalleled signature of the Majoranas -- is realized.

In this paper we consider a Kramers pair of Majoranas occurring in the presence of time-reversal symmetry in a Josephson junction on a 2d topological insulator, quantum spin Hall (QSH) system. We show that it changes the conductance at zero energy in a dramatic fashion, changing the system from having perfect normal transmission along the edge of the QSH system to having perfect Andreev transmission,\cite{Law09} for a wide range of parameters.
 We show how this changes at finite  temperature, and how to observe the proposed effect in a three-terminal setup. We start with a non-interacting fermions picture and then examine the effect of interactions on the system. The results prove to be quite different from the case of a single Majorana in contact with one \cite{Fid12} or two \cite{Aff13,Kom14} Luttinger liquids.

The rest of the paper is organized as follows: in this section we discuss  possible experimental setups and compare our model to the topological Kondo effect. In  Section \ref{sec:non-interacting} we study  details of the non-interacting problem, including  possible additional symmetries. In Section \ref{sec:interacting_small} we discuss the case of weak interactions, and in Section \ref{sec:interacting} we discuss the full solution of the interacting problem using the RG, how  additional symmetries can modify the RG flows and non-trivial critical points that can occur at strong coupling.
 In Section \ref{sec:conductance} we discuss the conductance predictions of our analysis. Finally, in Section \ref{sec:conclusion} we conclude and discuss  possible future research directions. A series of Appendices provide additional information. These include App. C which deals with the single-lead model, App. D which presents a tight-binding version of the model, which may be convenient for numerical studies, App. E which calculates the impurity entropy for various RG fixed points and verifies that it is consistent with the $g$-theorem governing impurity RG flows,\cite{Aff91} and App. F which derives the RG equations.

\subsection{Possible experimental setups}
\begin{figure}[t]
\includegraphics[width = \linewidth]{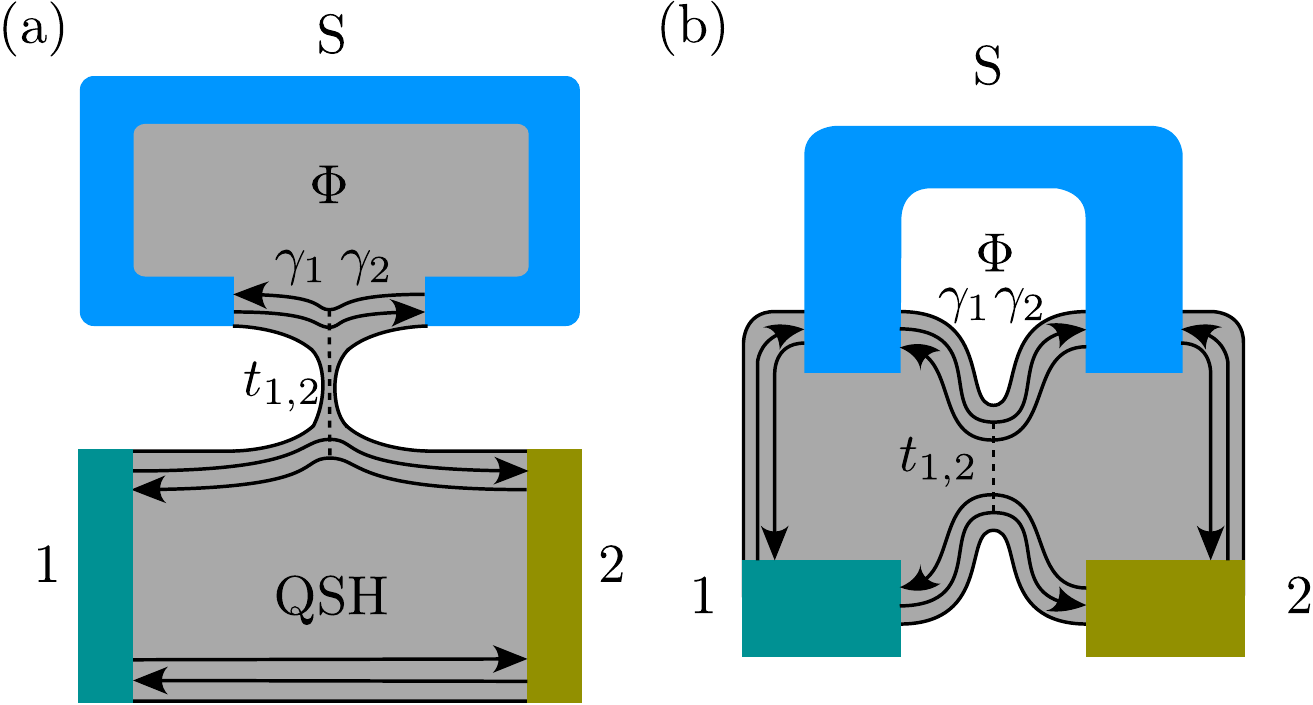}
\caption{Setups we use to study a Kramers pair of Majoranas in contact with a Luttinger liquid. The key ingredient of both the setups is the Josephson junction with phase difference $\pi$ on top of the quantum spin Hall bar. The junction hosts two Majorana bound states. The setups we consider has two normal and one superconducting lead. In either (a) or (b) there is a trivial conductance path: between left (1) and right (2) leads in (a) and between normal leads and superconducting one (SC) in (b). These conductance paths complicate but do not obscure the observations of the effect we propose.
}\label{fig:setup}
\end{figure}

We obtain the two Majoranas by inducing a phase difference $\pi$ Josephson junction on top of a 2d topological insulator exhibiting the QSH effect. 
See Figs. \ref{fig:setup}a) and \ref{fig:setup}b). 
It is straightforward to show the presence of two Majoranas in such junction, see Appendix (\ref{App:MBS}). 
Under time-reversal symmetry the phase difference across the junction changes sign; thus $\pi$ is the time-reversal invariant phase difference. Therefore, this situation corresponds to the symmetry class DIII with the two Majoranas in the Josephson junction. Time-reversal symmetry forbids the coupling of the two.

Notice that though we suggest magnetic field to create $\pi$-junction, the field can be made arbitrary small in the region of the junction. This can be done either by field focusing far away from the junction, or by making the superconducting loop arbitrary large and thus making the field to obtain $\pi$-junction arbitrary small.  This will make the effects of time-reversal symmetry breaking negligible. 

\subsection{Comparison with the topological Kondo effect}
A number of Majoranas coupled to a number of interacting or non-interacting channels has been considered by B\'eri and co-authors\cite{Ber12, Ber13, Alt14} in a series of papers. They dubbed the effect they observed ``Topological Kondo effect''. Due to the similarity of the setups it is instructive to point out the differences between our models.

First, there are obvious differences in the setups we consider. In our setup the superconductor is connected to an external lead, the charging energy is absent, and Andreev conductance is possible and instrumental for the effects we consider. In the topological Kondo effect setup the charging energy plays a  crucial role as the superconductor is small and floating. Due to the latter the Andreev conductance is always zero, while inter-channel conductance is present.
 In the case of the topological Kondo effect large charging energy projects the Hilbert space onto the states with a fixed number of particles on the superconducting island. This ensures that there are no direct tunneling terms in the Hamiltonian, while the co-tunneling terms are possible, 
see Eq. (2) of [\onlinecite{Ber12}]. In our case we study  Hamiltonians with  dominant tunneling terms,  with co-tunneling only emerging near the fixed point where the Majoranas are decoupled from the normal leads.

In summary though both in our case and in the case of topological Kondo effect the coupling of a few Majoranas to a few leads is considered, the differences are in the coupling of the superconductor to  an external lead which results in different dominant terms in the low-energy Hamiltonian, and in the presence of additional symmetries protecting the Majoranas from coupling. Our problem is a more direct generalization of the Fidkowski et al.\cite{Fid12} work, where Andreev conductance for a single Majorana bound state coupled to an interacting lead was considered.

\section{Non-interacting model\label{sec:non-interacting}}

We now consider the low-energy model of the system. For that we get rid of the gapped bulk degrees of freedom and concentrate on the edge of the QSHE which has 
right moving spin up particles and left moving spin down particles.  We start from phenomenologically introducing the low-energy field theory and discussing its symmetries.

The most general time-reversal invariant low energy non-interacting model, up to irrelevant operators,  coupling the QSHE edge with the Majoranas can be written as:
\bea
H &=& H_K + H_T + H_U,\label{eq:H} \nonumber \\
H_K &=& iv_F\int_{-\infty}^\infty dx [\psi_R^\dagger \partial_x\psi_R-\psi_L^\dagger \partial_x\psi_L],\nonumber \\
H_T &=& d^\dagger [t_1(\psi_R(0)+\psi_L^\dagger (0))+t_2(\psi_L(0)-\psi_R^\dagger (0)]\nonumber \\
&+&h.c.,\label{eq:H_T} \\
H_U &=& U_1 (\psi^\dag_R(0) \psi_R(0) + \psi^\dag_L(0) \psi_L(0)) \nonumber \\ 
&+& i U_2 (\psi^\dag_R(0) \psi^\dag_L(0) - \psi_L(0) \psi_R(0))\label{Hgen}
\eea
Here $d$ is a Dirac fermion composed of two Majoranas $\gamma_1$ and $\gamma_2$, $d=(\gamma_1+i\gamma_2)/2$; $\psi_{R, L}$ are right- and left-moving fermion fields in the low-energy field theory.  $H_K$ is the kinetic Hamiltonian, $H_T$ represents tunneling, and $H_U$ consists of perturbations that can be added at the origin, not breaking the time-reversal symmetry. To obtain this Hamiltonian we notice that the time-reversal symmetry is:
\begin{align}
\psi_R(x)&\to i\psi_L(x)\nonumber \\
\psi_L(x)&\to -i\psi_R(x)\nonumber \\
d&\to -id^\dagger \nonumber \\
i&\to -i.
\end{align}
Notice that the symmetry squares to $-1$ as it should for spinfull fermions,\cite{10foldway} even though we projected onto the low-energy effectively spinless model. To establish the form of the tunneling term $t_1$ in \eqref{eq:H_T} we notice that the $(t_1 d^\dag \psi_R(0) + t_1^* \psi_R(0)^\dag d)$ term under time-reversal symmetry goes to $(- t_1^* d \psi_L - t_1 \psi_L^\dag d^\dag)$. Analogously, one can check that other terms in \eqref{eq:H_T} and \eqref{Hgen} are allowed by time-reversal symmetry.

We can always make the tunneling amplitudes 
$t_1$ and $t_2$ real and positive, which we will henceforth assume,  by redefining the phases of the operators: 
\bea \psi_R(x)&\to& e^{i\alpha}\psi_R(x)\nonumber \\
 \psi_L(x)&\to& e^{-i\alpha}\psi_L(x)\nonumber \\
 d&\to& de^{i\beta}\eea
 for two independent phases $\alpha$ and $\beta$.

\subsection{$U(1)$ symmetric case}
We notice that the underlying physical Hamiltonian can have an additional spin-rotation $U(1)$ symmetry. It is approximately present in  real material if the effective spin-orbit interaction is close to zero near the junction. In other words, the symmetry is present if the spin quantization axis of the edge states does not vary spatially and there is no spin-flip in the constriction. If we assume the symmetry to be exact, we obtain a simplified and more tractable version of the problem, which we discuss in this section. 

We define the $U(1)$ symmetry by  $\psi_R \to e^{i\xi} \psi_R$,  $\psi_L \to e^{-i \xi} \psi_L$, and $d \to e^{i \xi} d$. Then only the $t_1$ term can 
appear in the  tunneling part of the Hamiltonian \eqref{eq:H}:
\begin{align}
H_T^0 = t d^\dagger (\psi_R(0)+\psi_L^\dagger (0))+h.c.
\end{align}
Alternatively we could require that  $d \to e^{-i \xi} d$ under the $U(1)$ symmetry, in which case only the $t_2$ term appears. Notice that $H_U$ preserve the symmetry.
The $U(1)$ symmetric Hamiltonian is equivalent to the well-known problem of a  quantum dot side-coupled to a quantum wire. To see the equivalence,
which will be  especially useful in the interacting case, we notice that   $\psi_L \to \tilde {\psi}^\dag_L$ transforms the Hamiltonian into a purely normal one:
\begin{align}
H_T \to t d^\dag (\tilde {\psi}_R(0) + \tilde {\psi}_L(0)) + h.c.\equiv \tilde H_T.\label{eq:HU(1)}
\end{align}
(We denote with the tilde operators in the transformed basis and refer to this as charge conjugation on left-movers or $C_L$ transformation.)

Let us now obtain the scattering matrix using the tunnelling Hamiltonian \eqref{eq:HU(1)}. We can write the eigenoperators as:
\be \tilde {\Gamma}_k\equiv \int_{-\infty}^\infty dx [\tilde {\psi}_R(x)\phi_R(x)+\tilde {\psi}_L(x)\phi_L(x)]+d\phi_d\ee
with $[\tilde {\Gamma}_k,\tilde {H}]=\epsilon_k \tilde {\Gamma}_k = v_Fk\tilde {\Gamma}_k$ 
the Schroedinger equations become:
\bea -iv\partial_x\phi_R+t\phi_d\delta (x)&=&v_Fk\phi_R(x)\nonumber \\
iv\partial_x\phi_L+t\phi_d\delta (x)&=&v_Fk\phi_L(x)\nonumber \\
t[\phi_R(0)+\phi_L(0)]&=&v_Fk\phi_d.\eea
We see that the solutions are step functions:
\bea \phi_R(x)&=&e^{ikx}R_+,\ \  (x>0)\nonumber \\
&=&e^{ikx}R_-,\ \  (x<0)\nonumber \\
\phi_L(x)&=&e^{-ikx}L_+,\ \  (x>0)\nonumber \\
&=&e^{-ikx}L_-,\ \  (x<0).\label{phi}\eea
The Schroedinger equations now reduce to:
\bea -iv_F(R_+-R_-)+t\phi_d&=&0\nonumber \\
-iv_F(L_--L_+)+t\phi_d&=&0\nonumber \\
t\sum_\pm (L_\pm +R_\pm )/2&=&v_Fk\phi_d.\eea
We can solve the last equation for $\phi_d$ and substitute into the first two equations.  These can be rewritten 
to express the outgoing amplitudes, $R_+$, $L_-$ in terms of the incoming amplitdues, $R_-$, $L_+$ in the form:
\be \left(\begin{array}{c} R_+\\L_-\end{array}\right)=\tilde {S}\left( \begin{array}{c} R_-\\L_+\end{array}\right)
\ee
where $\tilde {S}$ is the S-matrix in the transformed basis:
\be \tilde {S}=\left(\begin{array}{cc}-iv_F+u&u\\ u&-iv_F+u\end{array}\right)^{-1}\left(\begin{array}{cc}-iv_F-u&-u\\ -u&-iv_F-u\end{array}\right)
\ee
and 
\be u\equiv {t^2\over 2v_Fk}.\label{u}\ee
Solving:
\be \tilde {S}={1\over v_F+2iu}\left(\begin{array}{cc} v_F&-2iu\\-2iu&v_F\end{array}\right).\label{S}\ee
We label the S-matrix elements as
\be \tilde {S}=\left(\begin{array}{cc}\tilde {S}_{RR}&\tilde {S}_{RL}\\ \tilde {S}_{LR}&\tilde {S}_{LL}\end{array}\right).\ee
$\tilde {S}_{LR}$, for example, is the amplitude for an incoming right mover with spin up to turn into an outgoing left mover 
with spin down. 
At high enough energies, $u\ll v_F$, $|\tilde {S}_{LL}|\approx |\tilde {S}_{RR}|\approx 1$ and $|\tilde {S}_{LR}|\approx |\tilde {S}_{RL}|\approx 0$. 
But at low energies, this is reversed, corresponding to total reflection at zero energy. This anti-resonance is a consequence of the absence of a $d^\dagger d$ term 
in the Hamiltonian, due to time-reversal invariance.  The cross-over scale is $E\approx (t)^2/v_F$, quadratic in the tunneling amplitude $t$. This 
can be understood from the fact that $t$ has RG scaling dimension 1/2.

We now remember that initially the model had $\psi_L^\dag$ instead of $\tilde {\psi}_L$. In the original model the S-matrix is the same but now the interpretation of its elements change. $S_{LR}$ is now the amplitude
for an incoming right moving particle to turn into an outgoing left-moving hole and similarly for $S_{RL}$.  So, at zero energy we get perfect Andreev reflection, 
corresponding to $2e^2/h$ conductance. 

This simplified Hamiltonian is ideal to check the influence of additional possible terms in the non-interacting Hamiltonian, for example the $U_1$ term in Eq. (\ref{Hgen}).
In the particle-hole transformed model this term is:
\be \delta H=U[\tilde {\psi}^\dagger_R(0)\tilde {\psi}_R(0)-\tilde {\psi}^\dagger_L(0)\tilde {\psi}_L(0)]\ee
where we drop the subscript $1$ to simplify the notation. 
 Lets analyze its effects in the low energy normal dot model. The Schroedinger equations are modified to:
\bea -iv\partial_x\phi_R+t\phi_d\delta (x)+U\phi_R\delta (x)&=&v_Fk\phi_R(x)\nonumber \\
iv\partial_x\phi_L+t\phi_d\delta (x)-U\phi_L\delta (x)&=&v_Fk\phi_L(x)\nonumber \\
t[\phi_R(0)+\phi_L(0)]&=&v_Fk\phi_d.\eea
We make the same ansatz as above, Eq. (\ref{phi}), obtaining:
\bea -iv_F(R_+-R_-)+t\phi_d+U(R_++R_-)/2=0\nonumber \\
-iv_F(L_--L_+)+t\phi_d-U(L_-+L_+)/2=0\nonumber \\
t\sum_\pm (L_\pm +R_\pm )/2=v_Fk\phi_d.\eea
Note that, when $t=0$, the effect of $U$ is simply adding a transmission phase:
\bea R_+&=&{iv_F+U/2\over iv_F-U/2}R_-\nonumber \\
L_-&=&{iv_F-U/2\over iv_F+U/2}L_+.\label{TP}\eea
For non-zero $t$ 
\begin{align} S=&\left(\begin{array}{cc}-iv_F+u+U/2&u\\ u&-iv_F+u-U/2\end{array}\right)^{-1}\nonumber \\ \times&\left(\begin{array}{cc}-iv_F-u-U/2&-u\\ -u&-iv_F-u+U/2\end{array}\right)
\end{align}
where $u$ is defined in Eq. (\ref{u}). This is
\begin{align} S={-1 \over v_F(iv_F-2u)+U^2/4}\begin{pmatrix}(iv_F+U/2)^2&-2iv_Fu\\ -2iv_Fu&(iv_F-U/2)^2\end{pmatrix}
\end{align}
At $|u|\ll v_F$, $S$ becomes diagonal with elements given by Eq. (\ref{TP}). At low energies, where $|u|\gg v_F$, 
$S\to -\sigma_x$, as for $U=0$.  So, we again get an anti-resonance at zero energy with the crossover scale shifted to
\be v_F^2+U^2/4=v_Ft^2/E.\ee
Again in the spin-Hall system we get $2e^2/h$ conductance at zero temperature. 

\subsection{Case of broken $U(1)$ symmetry}
We now return to the generic non-interacting Hamiltonian of \eqref{eq:H}. We diagonalize,  $H\to\int{v_Fk\Gamma^\dag_k\Gamma\dn_k}$, where $\Gamma_k$'s are the scattering states:
\begin{align}
\Gamma_k=\phi_pd+\phi_h d^\dag + \int_{-\inf}^{\inf} dx\left[P_R(x)\psi_R(x)+H_R(x)\psi^\dag_R(x)\right.\nonumber\\ \left. +P_L(x)\psi_L(x)+H_L(x)\psi^\dag_L(x)\right]
\end{align}
We require $[\Gamma_k,H]=v_Fk\Gamma_k$, and use the commutation relations
\begin{align}
\com{\psi_R(x)}{H}=&iv_F\partial_x\psi_R(x)+t_2\delta(x)d^\dag+t_1\delta(x)d\nonumber \\ &+U_1\delta(x)\psi_R(0)+iU_2\delta(x)\psi^\dag_L(0)  \\
\com{\psi^\dag_R(x)}{H}=&iv_F\partial_x\psi^\dag_R(x)-t_1\delta(x)d^\dag-t^*_2\delta(x)d\nonumber \\ &-U_1\delta(x)\psi^\dag_R(0)+iU_2\delta(x)\psi_L(0)  \\
\com{\psi_L(x)}{H}=&-iv_F\partial_x\psi_L(x)-t_1\delta(x)d^\dag+t^*_2\delta(x)d\nonumber \\ &+U_1\delta(x)\psi_L(0)-iU_2\delta(x)\psi^\dag_R(0)  \\
\com{\psi^\dag_L(x)}{H}=&-iv_F\partial_x\psi^\dag_L(x)-t_2\delta(x)d^\dag+t^*_1\delta(x)d\nonumber \\ &-U_1\delta(x)\psi^\dag_L(0)-iU_2\delta(x)\psi_R(0) \\
\com{d}{H}=t_1\Big(\psi\dn_R(0&)+\psi^\dag_L(0)\Big)+t_2\Big(\psi\dn_L(0)-\psi^\dag_R(0)\Big)\\
\com{d^\dag}{H}=-t_1\Big(\psi&^\dag_R(0)+\psi\dn_L(0)\Big)-t_2\Big(\psi^\dag_L(0)-\psi\dn_R(0)\Big).
\end{align}
Inserting all these into $[\Gamma_k,H]$ and doing integration-by-parts we obtain
\begin{align}
v_FkP_R(x)=&-iv_F\partial_xP_R(x)+t_1\delta(x)\phi_p+t_2\delta(x)\phi_h \nonumber \\ &+U_1\delta(x)P_R(0)-iU_2\delta(x)H_L(0)\\
v_FkH_R(x)=&-iv_F\partial_xH_R(x)-t_2\delta(x)\phi_p-t_1\delta(x)\phi_h\nonumber \\ &-U_1\delta(x)H_R(0)-iU_2\delta(x)P_L(0)\\
v_FkP_L(x)=&iv_F\partial_xP_L(x)+t_2\delta(x)\phi_p-t_1\delta(x)\phi_h \nonumber \\&+ U_1\delta(x)P_L(0)+iU_2\delta(x)H_R(0)\\
v_FkH_L(x)=&iv_F\partial_xH_L(x)+t_1\delta(x)\phi_p-t_2\delta(x)\phi_h \nonumber \\ &- U_1\delta(x)H_L(0)+iU_2\delta(x)P_R(0)\\
v_Fk\phi_p=&t_1[P_R(0)+H_L(0)]+t_2[P_L(0)-H_R(0)]\\
v_Fk\phi_h=&-t_2[H_L(0)-P_R(0)]-t_1[P_L(0)+H_R(0)].
\end{align}
The $v_Fk$ term on the left, can be absorbed, with the ansatz:
\begin{align}
P_R(x)=e^{ikx}\tilde P_R(x), \; H_R(x)=e^{ikx}\tilde H_R(x),\nonumber \\  \qquad P_L(x)=e^{-ikx}\tilde P_L(x),\; H_L(x)=e^{-ikx}\tilde H_L(x).
\end{align}

The tilde functions, are just step functions, e.g. $\tilde P_R(x)=P_R(0^+)\Theta(x)+P_R(0^-)\Theta(-x)$ where $\Theta(x)$ is the Heaviside step function. Alternatively, we can write $\tilde P_R(x)=\bar P_R+\sgn(x)\delta P_R/2$, where $\bar P_R=[P_R(0^+)+P_R(0^-)]/2$ and $\delta P_R=P_R(0^+)-P_R(0^-)$.
So, we find
\bea
iv_F\delta P_R&=&U_1\bar P_R-iU_2\bar H_L+t_1\phi_p+t_2\phi_h\\
iv_F\delta H_R&=&-U_1\bar H_R-iU_2\bar P_L-t_2\phi_p-t_1\phi_h\\
-iv_F\delta P_L&=&U_1\bar P_L+iU_2\bar H_R+t_2\phi_p-t_1\phi_h\\
-iv_F\delta H_L&=&-U_1\bar H_L+iU_2\bar P_R+t_1\phi_p-t_2\phi_h\\
v_Fk\phi_p&=&t_1[\bar P_R+\bar H_L]+t_2[\bar P_L-\bar H_R]\\
v_Fk\phi_h&=&-t_2[\bar H_L-\bar P_R]-t_1[\bar P_L+\bar H_R]
\eea
Eliminating $\phi_p$ and $\phi_h$, we can write the equation in the form of 
\begin{align}
(iv_F\bb 1-A)\mat{P_R(0^+) \\ P_L(0^-) \\ H_R(0^+)  \\ H_L(0^-)}&=(iv_F\bb 1+A)\mat{P_R(0^-) \\ P_L(0^+)\\ H_R(0^-)  \\ H_L(0^+)} \nonumber \\ \so \mat{P_R(0^+) \\ P_L(0^-)\\ H_R(0^+)  \\ H_L(0^-)}&=S\mat{P_R(0^-) \\ P_L(0^+)\\  H_R(0^-)  \\ H_L(0^+)},
\end{align}
with 
\begin{align}
S=(iv_F\bb 1-A)^{-1}(iv_F\bb 1+A). \label{eq:S-matrix}
\end{align} 
$A=A^\dag$ ensures unitarity of the S-matrix.

The $S$-matrix elements can be labelled:
\begin{align}
S&=\mat{S^{ee} & S^{eh} \\ S^{he} & S^{hh}},\nonumber \\S^{xy}&=\mat{S_{RR} & S_{RL} \\ S_{LR} & S_{LL}}^{xy}, \qquad x,y=e,h.\label{SMlab}
\end{align}
\begin{widetext}
 We find
\be
A(\eps_k)=\frac{1}{2\eps_k}\begin{pmatrix}

U_1\eps_k+t_1^2+t_2t_2^2 & 0 & -2t_1t_2 & -iU_2\eps_k+{t_1^2-t_2^2} \\

0 & U_1\eps_k+t_1^2+t_2^2 & iU_2\eps_k+t_1^2-t_2^2  & 2t_1t_2\\

-2t_1t_2& -iU_2\eps_k+t_1^2-t_2^2& -U_1\eps_k+t_1^2+t_2^2 & 0 \\

iU_2\eps_k+t_1^2-t_2^2 & 2t_1t_2 & 0 & -U_1\eps_k+t_1^2+t_2^2

\end{pmatrix} \label{eq:A}
\ee
where 
\be \epsilon_k\equiv v_Fk.\ee
In the special case $U_1=U_2=0$ , we can find a simple expression for the $S$-matrix:
\be S={1\over 2t^2}\left(\begin{array}{cccc} (1+z)(t_1^2+t_2^2)&0&2(1-z)t_1t_2& (z-1)(t_1^2-t_2^2)\\
0&(1+z)(t_1^2+t_2^2)&(z-1)(t_1^2-t_2^2)&-2(1-z)t_1t_2\\
2(1-z)t_1t_2&(z-1)(t_1^2-t_2^2)&(1+z)(t_1^2+t_2^2)&0\\
(z-1)(t_1^2-t_2^2)&-2(1-z)t_1t_2&0& (1+z)(t_1^2+t_2^2)
\end{array}\right)\label{SU0}
\ee
\end{widetext}
where 
\be z\equiv {iv_F-2t^2/(v_Fk)\over iv_F+2t^2/(v_Fk)}.\label{eqz}\ee
and
\begin{align}
t = \sqrt{t_1^2 + t_2^2}.
\end{align}
The zero elements in $S$, signifying the absence of normal scattering, are a consequence of time-reversal symmetry.  Note the absence 
of Andreev transmission when either $t_1$ or $t_2=0$, as required by $U(1)$ symmetry. More surprisingly, there is an absence of Andreev reflection
when $t_1=t_2$. 
This is a consequence of a peculiar $U(1)$ symmetry which occurs in that case.  From Eq. (\ref{Hgen}) we see that, when $t_1=t_2$ 
the Majorana modes couple to a Dirac fermion:
\be \psi_0\equiv [(\psi_L+\psi_L^\dagger)+(\psi_R-\psi_R^\dagger )]/2.\ee
The Hermitean part of $\psi_0$ is the Hermitean part of $\psi_L(0)$ and the anti-Hermitean part of $\psi_0$ is the anti-Hermitean part of $\psi_R(0)$. 
The full Hamiltonian has a $U(1)$ symmetry that rotates the phase of $\psi_0$ and the phase of $d$ oppositely. This corresponds to an $O(2)$ 
symmetry mixing the Hermitean part of $\psi_L(x)$ with the anti-Hermitean part of $\psi_R(-x)$. 

To prove that the symmetry forces Andreev transmission let us decompose $\psi_{R/L}$ into Majorana components:
\be \psi_{R/L}(x)=\chi_{R/L}+i\chi_{R/L}'.\ee
Then the special $U(1)$ symmetry when $t_1=t_2$ is:
\be \left(\begin{array}{c} \chi_L\\ \chi_R'\end{array}\right)\to 
{\cal M}\left(\begin{array}{c} \chi_L\\ \chi_R'\end{array}\right)\ee
where $M$ is an $SO(2)$ matrix. $\chi_L'$ and $\chi_R$ are unaffected by this rotation. Andreev transmission 
corresponds to
\be \psi_{R/L}(0^-)=\pm \psi_{R/L}^\dagger (0^+).\ee
This implies
\be  \left(\begin{array}{c} \chi_L\\ \chi_R'\end{array}\right)(0^-)=- \left(\begin{array}{c} \chi_L\\ \chi_R'\end{array}\right)(0^+),\ee
which is consistent with the $O(2)$ symmetry. On the other hand, Andreev reflection implies
\be \psi_R(0^\pm )=\psi_L^\dagger (0^\pm )\ee
and hence 
\be \left(\begin{array}{c} \chi_L\\ \chi_R'\end{array}\right)(0^\pm )=
\left(\begin{array}{c} \chi_R\\ -\chi_L'\end{array}\right)(0^\pm ).\ee
This is {\it not} consistent with the $O(2)$ symmetry under which the left-hand vector rotates but the right-hand vector does not.  Thus the symmetry can only support Andreev transmission.

Unlike the $U(1)$ symmetry which occurs when $t_1$ or $t_2=0$, this $U(1)$ symmetry occurring when $t_1=t_2$ is not present once interactions are included since it is non-local.

From Eq. (\ref{eq:A})  it is clear that the $H_U$ terms drop out for $\eps_k\to 0$. So, we see although the $H_U$ terms  
are marginal at the high energy normal transmission fixed point, at the infrared (IR) fixed point, they are irrelevant. 
\begin{widetext}

\subsubsection{$\epsilon \to 0$}
For small energies $\epsilon\to 0$ we have $z\to-1$ and Eq.\,\pref{SU0} simplifies to:
\be S\to {1\over 2t^2}\left(\begin{array}{cccc} 0&0&4t_1t_2&-2(t_1^2-t_2^2)\\ 
0&0&-2(t_1^2-t_2^2)&-4t_1t_2\\ 
4t_1t_2&-2(t_1^2-t_2^2)&0&0\\ 
-2(t_1^2-t_2^2)&-4t_1t_2&0&0\end{array}\right).\label{eq:S}\ee
Note that purely Andreev processes occur at zero energy: both Andreev scattering and Andreev transmission. In the $U(1)$ symmetric case, 
where either $t_1$ or $t_2=0$, we get purely Andreev reflection at zero energy. 

\subsubsection{$t_1=t_2$ and $U_2=0$}
We see that for $t_1=t_2=t/\sqrt{2}$, and $U_2=0$ the blocks of the A-matrix are all diagonal:
\be
A(\eps_k)=\frac{1}{2\eps_k}\begin{pmatrix}

U_1\eps_k+t^2 & 0 & -t^2& 0 \\
0 & U_1\eps_k+t^2 & 0&t^2\\
-t^2&0 & -U_1\eps_k+t^2& 0 \\
0& t^2 & 0&-U_1\eps_k+t^2
\end{pmatrix} \label{eq:AT}
\ee
From this, we get the S-matrix:
\begin{align}
S(\epsilon) = \frac{1}{\epsilon(U_1^2 + 4 v_F^2) + 4 i v_F t^2} \begin{pmatrix}
- \epsilon(U_1 + 2 i v_F)^2&0&4 i v_F t^2&0\\
0&- \epsilon(U_1 + 2 i v_F)^2&0&-4 i v_F t^2\\
4 i v_F t^2&0&- \epsilon(U_1 - 2 i v_F)^2&0\\
0&-4 i v_F t^2&0&- \epsilon(U_1 - 2 i v_F)^2
\end{pmatrix}
\end{align}

For $\eps\to 0$ we get $S_{RR}=S_{LL}=\tau_x$ -- pure Andreev transmission. An incoming electron  $P_R(0^-)$ becomes a reflected hole  $H_R(0^+)$ and two electrons are transferred to the superconductor (one from contact 1 and one from contact 2).
\end{widetext}

\section{Interacting model\label{sec:interacting_small}}
We now extend the previous analysis to include interactions, both global along the spin-Hall edges and local at the junction 
with the Majorana modes. We model the important edge interactions by
\be H_{\hbox{int}}=V\int_{-\infty}^\infty dx :\psi^\dagger_R\psi_R::\psi^\dagger_L\psi_L:\label{eqHV}\ee
where the double dots denote normal ordering. 
Note that, appropriate to the set-ups of Fig. \ref{fig:setup}, we have labeled the right and left side of 
the junction by $x>0$ and $x<0$ respectively and considered the case where the contacts are infinitely far away. 
We will later consider the 
effect of a finite distance to the contacts.  
It will sometimes be convenient below to regard the $x>0$ and $x<0$ regions as two different leads, in which 
case we introduce two different fields $\psi_{L/R,1}$, $\psi_{R/L,2}$ both defined on the $x>0$ axis. 

To study the low-energy properties of the model we bosonize the Luttinger liquid. We use the notations:
\begin{align}
\psi_{L/R} \propto \Gamma \exp\left\{i \sqrt{\pi}[\phi (x) \pm \theta (x)\right\},\label{bosonize}
\end{align}
where $\Gamma$ is a Klein factor, and $\phi$ and $\theta$ are the usual boson fields with the commutation relation $[\phi (x), \theta (y)] = - i \Theta(y-x)$. We explicitly obtain these bosonisation formulas in Appendix A. The time-reversal symmetry is then:
\begin{align}
\phi &\to - \phi, \\
\theta &\to \theta + \sqrt{\pi}/2, \\
i &\to -i.
\end{align}
and the $U(1)$ symmetry is:
\begin{align}
\theta \to \theta - \frac{\xi}{\sqrt{\pi}}.
\end{align}
The Hamiltonian in the bulk of the system is:
\begin{align}
H = \frac{1}{2}u\int_{-\infty}^\infty dx \left[K \left(\frac{\partial\phi}{\partial x}\right)^2+K ^{-1} \left(\frac{\partial\theta}{\partial x}\right)^2\right].\label{eq:H_bosonized}
\end{align}
$K$ and $u$ are expressed via the parameters of the fermionic model:
\begin{align}
u = \sqrt{v_F^2 - \frac{V^2}{16\pi^2}}, \; K = \sqrt{\frac{v_F - \frac{V}{4\pi}}{v_F + \frac{V}{4\pi}}}.
\end{align}

Besides the tunneling terms present in the non-interacting case, $t_1$ and $t_2$, there are the following four-fermion boundary terms:
\begin{align}
&U_{NT} (d^\dag d - 1/2) (\psi^\dag_R \psi_R - \psi^\dag_L \psi_L)\nonumber \\ &\propto U_{NT} (d^\dag d - 1/2) \partial_x \phi(0),\label{eq:U_nt}\\
&U_{AR} (d^\dag d - 1/2) (\psi^\dag_R \psi^\dag_L + \psi_L \psi_R)\nonumber \\ 
&\propto U_{AR} (d^\dag d - 1/2) \sin 2 \sqrt{\pi} \phi(0), \label{eq:U_ar}\\
&U_{NR} (d^\dag d - 1/2) (\psi^\dag_R \psi_L + \psi^\dag_L \psi_R) \nonumber\\ &\propto |U_{NR}| (d^\dag d - 1/2)
\cos (2\sqrt{\pi}\theta(0)+\alpha_{NR}),\label{eq:Jz}
\end{align}
Here $\alpha_{NR}$ is the phase of $U_{NR}$. We will set them to zero in most of the cases, besides at NR fixed point, where they become important. 

The three terms \eqref{eq:U_nt}-\eqref{eq:Jz} favor Normal Transmission, Andreev Reflection, and single-particle Normal Reflection. While the first 
two respect the $U(1)$ symmetry introduced in Sec. II, $U_{NR}$ does not. 

Additionally, we have the terms discussed in the non-interacting case, $U_1$ and $U_2$, \eqref{Hgen}. They bosonize to:
\begin{align}
&U_1 \partial_x \theta(0), \\
&U_2 \cos 2\sqrt{\pi} \phi(0).
\end{align}
$U_1$ can be included into the Hamiltonian by a local unitary transformation (the like of which, removing $U_{NT}$ from the Hamiltonian will be discussed in the next section). $U_2$ is not generated in RG up to second order in $t$, as we show in the Appendix F, and is irrelevant. Therefore even if it is present in the Hamiltonian it simply renormalizes to zero. We do not consider these terms below.

We now proceed with studying the model close to the non-interacting point, where $K>1/2$ and all local couplings,  tunneling, $U_{NT}$, and one of $U_{AR}$ and $U_{NR}$ are small. The analysis of the present section will be most relevant for the experiments. In the next section we will treat different parameter ranges, including arbitrary $K<1$ and $U_{NT}$. We note that $U_{NR}$ is the only term which may be not small in the experimental system, since it results from direct Coulomb interaction between the localized level and the bulk fermions. If it is large, this does not change the analysis below besides moving the starting point of RG towards NR critical point. Even if small in the real system, we do not need to assume $U_{AR}$ is small. When it is large, the RG will start from the vicinity of the AR fixed point. We assume however that the product of $U_{AR}U_{NT}$ is small so that there is no frustration when the RG starts.

\subsection{Interacting $U(1)$ symmetric  case}

\subsubsection{NT critical point}

We notice that the dimension of the tunneling operators at NT critical point is $(K+1/K)/4$. The dimension is smaller than $1$ for any $1/2<K<1$, therefore the operator is relevant for not too strong interactions. This means that the high-energy NT critical point is unstable to tunneling to Majoranas. This is consistent with the non-interacting results from above, where the normal transmission at high energies was replaced by Andreev reflection at zero energy for the $U(1)$ symmetric case.

\subsubsection{Stability of AR critical point}

We use the $C_L$ transformed fields defined in Sec. IIA. This brings our Hamiltonian to the familiar form of the side-coupled normal quantum 
dot\cite{Gol10} and causes a change in the sign of the interaction term in Eq.\,\pref{eqHV} so that the corresponding Luttinger parameter is $\tilde  K\approx 1/K$ for $K\approx 1$. After this transformation the time-reversal symmetry becomes:
\bea \tilde {\psi}_L&\to& \tilde {\psi}_R^\dagger \nonumber \\
\tilde {\psi}_R&\to&\tilde {\psi}_L^\dagger \nonumber \\
d&\to&-d^\dagger \nonumber \\
i&\to& -i.\label{TRGB}
\eea
Notice that it does not anymore square to $-1$, but to $1$. We denote this symmetry CT for the reasons discussed below. This is our choice, and we have it due to the $U(1)$ symmetry allowing for block-diagonalizing of the Hamiltonian and choosing arbitrary sign of square of the time-reversal symmetry acting between the Hamiltonian blocks\cite{10foldway}. We may identify this symmetry as a product of time reversal symmetry
\bea\tilde {\psi}_R(x)&\leftrightarrow& \tilde {\psi}_L(x)\nonumber \\
d&\to& d\nonumber \\
i&\to &-i\label{TR3}
\eea
and charge conjugation:
\bea \tilde {\psi}_{L/R}(x)&\to& \tilde {\psi}_{L/R}^\dagger (x)\nonumber \\
d&\to& -d^\dagger \nonumber \\
i&\to& i.\eea
(We dropped the factors of $i$ from time-reversal for convenience; this is only possible in the $U(1)$ symmetric case.) 

 In the resonant case, with no $d^\dagger d$ term, which follows from time-reversal symmetry, 
it was concluded in [\onlinecite{Gol10}]  that there is a zero transmission or 
perfect normal reflection fixed point for $\tilde  K>1$ which is the case of interest for us.  After undoing the $C_L$-transformation, this  becomes Andreev reflection in the original model.

Lets check whether this normal reflection fixed point is stable. It corresponds to the boundary conditions:
\be \tilde {\psi}_R(0^\pm)=-\tilde {\psi}_L(0^\pm ).\label{bc}\ee
Any relevant interaction which couples the 2 sides together would destabilize this fixed point.  
With $U(1)$ symmetry the only candidates are
constructed from $\tilde {\psi}_R^\dagger (0^+)\tilde {\psi}_L(0^-)$ and related terms.  Note that under CT symmetry:
\be \tilde {\psi}_R^\dagger (0^+)\tilde {\psi}_L(0^-)\to \tilde {\psi}_L(0^+)\tilde {\psi}_R^\dagger (0^-)=-\tilde {\psi}_R^\dagger (0^-)\tilde {\psi}_L(0^+).\ee
Thus there are 2 possibilities for relevant CT invariant Hermitean terms in the effective Hamiltonian. These are:
\bea \delta H_1&=&U_1[ \tilde {\psi}_R^\dagger (0^+)\tilde {\psi}_L(0^-)-\tilde {\psi}_R^\dagger (0^-)\tilde {\psi}_L(0^+)\nonumber \\ &&+\tilde {\psi}_L^\dagger (0^-)\tilde {\psi}_R(0^+)
-\tilde {\psi}_L^\dagger (0^+)\tilde {\psi}_R(0^-)]\nonumber \\
\delta H_2&=&iU_2[ \tilde {\psi}_R^\dagger (0^+)\tilde {\psi}_L(0^-)+\tilde {\psi}_R^\dagger (0^-)\tilde {\psi}_L(0^+)\nonumber \\ &&-\tilde {\psi}_L^\dagger (0^-)\tilde {\psi}_R(0^+)
-\tilde {\psi}_L^\dagger (0^+)\tilde {\psi}_R(0^-)]
\eea
where $U_1$ and $U_2$ are real.  (Note that no terms involving $d$ can appear at the perfect reflection fixed point, since $d$ is screened there.) 
After imposing the boundary conditions of Eq. (\ref{bc}) these become:
\bea \delta H_1&=&U_1[- \tilde {\psi}_R^\dagger (0^+)\tilde {\psi}_R(0^-)+\tilde {\psi}_R^\dagger (0^-)\tilde {\psi}_R(0^+)\nonumber \\ &&-\tilde {\psi}_R^\dagger (0^-)\tilde {\psi}_R(0^+)
+\tilde {\psi}_R^\dagger (0^+)\tilde {\psi}_R(0^-)]\nonumber \\
\delta H_2&=&iU_2[ -\tilde {\psi}_R^\dagger (0^+)\tilde {\psi}_R(0^-) -\tilde {\psi}_R^\dagger (0^-)\tilde {\psi}_R(0^+)\nonumber \\ &&+ \tilde {\psi}_R^\dagger (0^-)\tilde {\psi}_R(0^+)
+\tilde {\psi}_R^\dagger (0^+)\tilde {\psi}_R(0^-)]
\eea
which are both zero.  Thus we conclude that CT symmetry stabilizes the perfect reflection fixed point in the $C_L$-transformed model. This implies 
that time-reversal stabilizes the perfect Andreev reflection fixed point in the original model. 

Thus we conclude that for small enough interaction strength the RG flow goes from NT to AR fixed points. This is consistent with the non-interacting results obtained above, since there at high energies the conductance correspond to the normal transmission, while at zero energy the incoming electrons always undergo Andreev reflection. Let us now turn to the effects of the $U(1)$ symmetry breaking. 

\subsection{Interacting $U(1)$ breaking case}

\subsubsection{NT critical point}
Here we reiterate that the NT critical point is unstable to the tunneling terms \eqref{eq:H_T}. Besides, the dimension of $U_{NR}$ (allowed when $U(1)$ is broken) is $K$, therefore it is also relevant at the NT critical point. $U_{NR}$ has the form of the usual Coulomb interaction and therefore we cannot assume that it is small in the experiment. Depending on the actual size of $U_{NR}$ we start near the NT critical point (small $U_{NR}$), or near the NR critical point (large $U_{NR}$). We will discuss the NR critical point below.

\subsubsection{AR and AT critical points}

We reinterpret the non-interacting results in the $U(1)$ symmetry breaking case as follows. We know that the scattering matrix at high energies shows dominating normal transmission (NT). Therefore, the ultraviolet fixed point is the NT one. At zero energy no normal transmission is possible. Andreev process -- reflection (AR), transmission (AT), or anything in between can happen. We interpret this result as the system having a line of fixed points, connecting the AR and AT fixed points. As we see below, this is a pathological situation, destabilized by arbitrary weak interactions. In particular, AR becomes unstable and AT becomes the absolutely stable fixed point. 
Therefore the line collapses onto one point, AT.

We start our argument by considering the stability of the Andreev reflection fixed point when we break the $U(1)$ symmetry. It is now important to use the original 
definition of time-reversal in Eq. (\ref{TR3}) since the simplified version with factors of $i$ dropped is not a symmetry when both $t_1$ and $t_2$ are non-zero. After the $C_L$-transformation the time-reversal symmetry becomes:
\bea \tilde {\psi}_R&\to& i\tilde {\psi}_L^\dagger \nonumber \\
 \tilde {\psi}_R^\dagger&\to&- i\tilde {\psi}_L \nonumber \\
\tilde {\psi}_L^\dagger&\to&-i\tilde {\psi}_R \nonumber \\
\tilde {\psi}_L&\to&i\tilde {\psi}_R^\dagger  \nonumber \\
d&\to&-id^\dagger \nonumber \\
i&\to& -i.\label{CT}\eea
Possible $U(1)$ breaking terms 
are $\tilde {\psi}_R(0^+)\tilde {\psi}_L(0^-)$ and related terms.  The time-reversal symmetry maps:
\be \tilde {\psi}_R(0^+)\tilde {\psi}_L(0^-)\to -\tilde {\psi}_L^\dagger (0^+)\tilde {\psi}_R^\dagger (0^-).\ee
Again there are 2 possible time-reversal invariant Hermitean terms:
\bea \delta H_3&=& U_3[\tilde {\psi}_R(0^+)\tilde {\psi}_L(0^-)- \tilde {\psi}_L^\dagger (0^+)\tilde {\psi}_R^\dagger (0^-)\nonumber \\ &&+\tilde {\psi}_L^\dagger (0^-)\tilde {\psi}_R^\dagger (0^+)
-\tilde {\psi}_R(0^-)\tilde {\psi}_L(0^+)]\nonumber \\
 \delta H_4&=& iU_4[\tilde {\psi}_R(0^+)\tilde {\psi}_L(0^-)+ \tilde {\psi}_L^\dagger (0^+)\tilde {\psi}_R^\dagger (0^-)\nonumber \\ &&-\tilde {\psi}_L^\dagger (0^-)\tilde {\psi}_R^\dagger (0^+)
-\tilde {\psi}_R(0^-)\tilde {\psi}_L(0^+)]
\eea
for real $U_3$ and $U_4$.   Imposing the boundary conditions of Eq. (\ref{bc}), these become:
\bea \delta H_3&=& U_3[-\tilde {\psi}_R(0^+)\tilde {\psi}_R(0^-)+ \tilde {\psi}_R^\dagger (0^+)\tilde {\psi}_R^\dagger (0^-)\nonumber \\ &&-\tilde {\psi}_R^\dagger (0^-)\tilde {\psi}_R^\dagger (0^+)
+\tilde {\psi}_R(0^-)\tilde {\psi}_R(0^+)]\nonumber \\
 \delta H_4&=& iU_4[-\tilde {\psi}_R(0^+)\tilde {\psi}_R(0^-)- \tilde {\psi}_R^\dagger (0^+)\tilde {\psi}_R^\dagger (0^-)\nonumber \\ &&+\tilde {\psi}_R^\dagger (0^-)\tilde {\psi}_R^\dagger (0^+)
+\tilde {\psi}_R(0^-)\tilde {\psi}_R(0^+)]
\eea
We see that in this case the identical terms add instead of canceling each other. Therefore the tunneling terms are allowed by symmetry. 
The normal reflection boundary condition in the $C_L$-transformed model, Eq. (\ref{bc}), imply $\tilde  \theta (0)=$ constant. Applying the bosonization formulas of 
Eq. (\ref{bosonize}) in terms of bosons $\tilde  \phi$, $\tilde  \theta$, $\tilde  \psi_R(0^+)\tilde  \psi_R(0^-)$ 
 bosonizes $\propto\exp [i\sqrt{\pi}[\tilde  \phi (0^+)]\exp [i\sqrt{\pi}\tilde  \phi (0^-)]$ of dimension $1/\tilde  K$. Since we consider $\tilde  K>1$ in the 
$C_L$-transformed model these are relevant. Here we use the fact that boundary operators of dimension $<1$ are relevant. 
So, we conclude that the perfect normal reflection fixed point is unstable in the $C_L$-transformed model once we 
break $U(1)$ symmetry. Analogously, the perfect Andreev reflection fixed point is unstable in the original model once we break $U(1)$.

Now lets consider the stability of the Andreev transmission fixed point in the interacting case. The corresponding boundary conditions are (we now work with the original fermions, 
not making the $C_L$ transformation):
\bea \psi_R(0^-)&=&-\psi_R^\dagger (0^+)\nonumber \\
\psi_L(0^-)&=&\psi_L^\dagger (0^+)\label{bcFAT}
\eea
up to some phases that do not change the physics, since they can always be incorporated into redefinition of the bosonic fields. It is convenient to fold the system so that it is defined at $x>0$ only with 2 channels of left and right movers. For $x>0$:
\bea \psi_R(x)&\equiv& \psi_{R1}(x)\nonumber \\
\psi_L(x)&\equiv& \psi_{L1}(x) \nonumber \\
\psi_R(-x)&\equiv& \psi_{L2}(x) \nonumber \\
\psi_L(-x)&\equiv& \psi_{R2}(x)\
\eea
Note that, with these definitions, we have 2 channels of left and right movers at $x>0$ with only intra-channel interactions (assuming the 
initial interactions are short range). The Andreev transmission boundary conditions become:
\bea -\psi_{L2}(0)&=&\psi_{R1}^\dagger (0)\nonumber\\
 \psi_{R2}(0)&=&\psi_{L1}^\dagger (0).\label{eq:AT_fermion}
\eea
In the folded formulation these look like cross-channel Andreev reflection boundary conditions. This is similar to the usual s-wave Andreev reflection in a semi-infinite wire. After bosonization the Andreev transmission boundary conditions read (two different solutions for bosons are possible):
\bea \phi_2(0)+\theta_2(0)&\stackrel{{\rm mod}\, 2\sqrt{\pi}}{=}&-\phi_1(0)+\theta_1(0)+\sqrt{\pi} \nonumber \\
\phi_2(0)-\theta_2(0)&\stackrel{{\rm mod}\, 2\sqrt{\pi}}{=}&-\phi_1(0)-\theta_1(0).
\eea
These imply
\be \sqrt{\pi} /2 \stackrel{{\rm mod}\, \sqrt{\pi}}{=}\phi_1(0)+\phi_2(0) = -[\theta_1(0)-\theta_2(0)]\ee
It is convenient to switch to linear combination of boson fields, each of which has $K<1$ for the repulsive interactions:
\bea \phi_\pm &\equiv& {\phi_1\pm \phi_2\over \sqrt{2}}\nonumber \\
\theta_\pm &\equiv& {\theta_1\pm \theta_2\over \sqrt{2}}
\eea
Thus the Andreev transmission boundary conditions become:
\be \phi_+(0)=-\theta_-(0)\stackrel{{\rm mod}\, \sqrt{\pi/2}}{=}\sqrt{\pi /8}.\label{ATbc}\ee
Now lets consider the 3 types of processes that might destabilize the fixed point: normal reflection, Andreev reflection, normal transmission, and 
see if any corresponding interactions are allowed by the time-reversal symmetry and are relevant. 
\subsubsection*{Normal reflection}
Single-particle normal reflection enters as an operator:
\be \psi^\dagger_R(0^+)\psi_L(0^+)\ee
and related terms.  Under the CT transformation:
\begin{align}
 u \psi^\dagger_R(0^+)\psi_L(0^+)&\to - u^* \psi_L^\dag(0^+)\psi_R (0^+)\nonumber \\ &=-(u \psi^\dagger_R(0^+)\psi_L(0^+))^\dag.
\end{align}
Thus this term breaks time-reversal symmetry. This is in accordance with the common wisdom about the QSH edge, where no elastic backscattering can be induced in presence of time-reversal symmetry.

\subsubsection*{Andreev reflection}
This corresponds to terms like
\be \psi_R(0^+)\psi_L(0^+).\ee
This transforms under time-reversal symmetry to
\be -\psi_R (0^+)\psi_L (0^+).\ee
Therefore a term 
\begin{align}
i U (\psi_R(0^+) \psi_L(0^+) - h.c.)
\end{align}
can be present in the Hamiltonian. This bosonize to:
\begin{align}
i U e^{2 i \sqrt{\pi} \phi_1}+h.c.=iUe^{i\sqrt{2\pi}(\phi_++\phi_-)}+h.c.\nonumber \\ \to - 2U \cos \sqrt{2\pi}\phi_-
\end{align}
where the Andreev transmission boundary conditions of Eq. (\ref{ATbc}) were used. This operator has 
 dimension $1/K$. Therefore this term is irrelevant for $K<1$. For $K=1$ the term is marginal and it tunes between the AT and AR fixed points.

\subsubsection*{Normal Transmission}

\begin{figure}[t]
\includegraphics[width=0.6\linewidth]{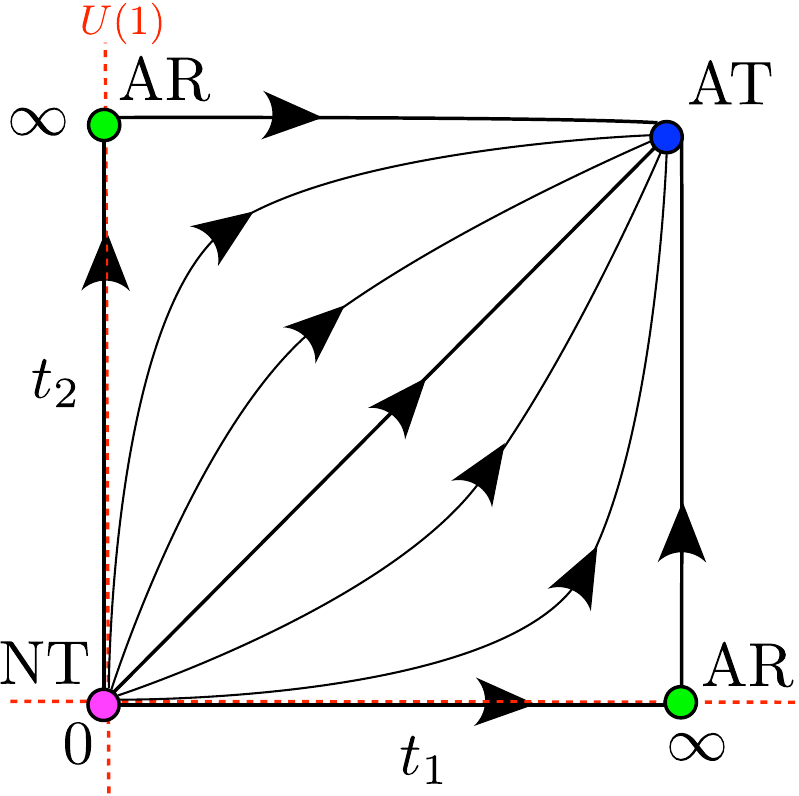}
\caption{RG flow for the model for $K>1/2$ and small initial $\tilde{U}_{NT}$ (region I from Fig. \ref{fig:regions}) projected onto the $t_1$-$t_2$ coordinate space. Fixed points present are: normal transmission (NT), corresponding to no tunneling into Majoranas, Andreev reflection (AR), corresponding to $t_1=0$ or $t_2=0$ and the other one renormalized to $\infty$, and Andreev transmission (AT), corresponding to $t_1\to \infty$ and $(t_1 - t_2)/t_1\to 0$. The RG flow between them is shown by the arrows. AR fixed point is stable if $U(1)$ symmetry is preserved, while AT fixed point is the stable one if the symmetry is broken.}\label{fig:RG_fixed_points}
\end{figure}

This corresponds to the likes of
\be \psi^\dagger_R(0^-)\psi_R(0^+).\ee
Under CT this goes to
\be \psi_L^\dag(0^-)\psi_L (0^+).\ee
So a term
\be \psi_R^\dagger (0^-)\psi_R(0^+)+ \psi_L^\dagger (0^-)\psi_L(0^+)+h.c.\ee
is allowed by time-reversal symmetry.  In folded notation the first term is:
\be \psi_{L2}^\dagger (0)\psi_{R1}(0).\ee
The Andreev transmission boundary conditions of Eq. (\ref{eq:AT_fermion}) allow us to write this as
\be \psi_{R1}(0)\partial_x\psi_{R1}(0)\ee
where the derivative is required due to Fermi statistics. We see that this has dimension $2$ for free fermions, $K=1$. In bosonized form, 
\begin{align}\psi_{L2}^\dagger (0)\psi_{R1}(0)&\propto \exp [i\sqrt{\pi}(-\phi_2-\theta_2+\phi_1-\theta_1)]\nonumber \\ &=\exp [i\sqrt{2\pi}(-\phi_--\theta_+)]
\end{align}
Neither of these bosons is pinned and the total dimension is
\be d=(K+1/K)>2.\label{eq:dimension_normal_transmission}\ee
consistent with $d=2$ for the non-interacting case. 
So, normal transmission is strongly irrelevant at the Andreev transmission fixed point for any $K$ and the leading irrelevant process is Andreev reflection. 
Andreev transmission is thus the stable fixed point of our model. 

The RG flow between the fixed points of the model for small enough interactions on the edge is shown in detail in Fig. \ref{fig:RG_fixed_points}. 

\subsubsection{NR fixed point}

Finally, if $U_{NR}$ grows to infinity or is very large to start with, $(d^\dag d - 1/2) \cos (2\sqrt{\pi}\theta(0)+\alpha_{NR})$ is pinned. This has the following minima:
\begin{align}
\theta+\frac{\alpha_{NR}}{2\sqrt{\pi}} &= 0, \sqrt{\pi}; \; d^\dag d = 0;\\
\theta+\frac{\alpha_{NR}}{2\sqrt{\pi}} &= \pm \sqrt{\pi}/2; \; d^\dag d = 1.
\end{align}
 Time reversal symmetry is spontaneously 
broken at these critical points, allowing normal scattering, with its phase depending on $\langle d^\dagger d-1/2\rangle $ and $\alpha_{NR}$. For each choice of $\langle d^\dagger d-1/2\rangle$ despite both values of possible $\theta$ pinning correspond to the same boundary condition on the fermion field (normal reflection with a phase which depends on $\alpha_{NR}$). Thus the choice of $\langle d^\dagger d-1/2\rangle$ determines the ground state uniquely at the fixed point. The ground state degeneracy is therefore $2$, corresponding to the two possible occupations of the $d$ fermion. Importantly, operator $d$ is not screened at this fixed point, since the two ground states correspond to different values of occupation number $d^\dag d$. Let us now analyze the stability of this critical point.

The most relevant operator that can destabilize the fixed point is tunneling, $d \psi^\dag_{R}$ \textit{et cetera}, which have the scaling dimension $1/(2 K)$. 
For $K>1/2$ these terms are relevant, making the fixed point unstable and presumably leading to a  flow to the Andreev transmission fixed point, the only stable fixed point in this range of $K$ for the broken $U(1)$ symmetry case.  We note that when 
$t_1=t_2=0$ but $U_{NR}\neq 0$ there is a different kind of $U(1)$ symmetry, broken only by the irrelevant 
$U_{AR}$ (and $U_2$) terms.  This simply transforms $\psi_{R/L}\to e^{i\alpha}\psi_{R/L}$, corresponding to 
ordinary charge conservation. 

Notice that the $\tilde{U}_{NT}$ term is absent at this fixed point: though $d$ is not fixed, $\partial_t \theta$ is fixed at this fixed point, therefore forbidding $\partial_x \phi$ entering the $\tilde{U}_{NT}$ term. Next, the term $\partial_x\theta^{\pm}$ is allowed at the fixed point and it just shifts the phase of backscattering. 

In the next section we discuss the general phase diagram of the model, and in Fig. \ref{fig:regions} the flow in Fig. \ref{fig:RG_fixed_points} corresponds to the region I. We have checked the consistency of the RG flow with the g theorem in  Appendix \ref{app:Z}.

\section{Full phase diagram of the interacting model\label{sec:interacting}}

Before proceeding with the discussion of the full phase diagram of the interacting model for all values of (repulsive) interaction strength, we notice that there are many other couplings that are irrelevant at $K=1$, but may become relevant for strong interactions. The most relevant of them for all $K$'s is:
\begin{align}
&U_{2NR} \psi_R^\dag(0^+) \partial_x \psi_R^\dag(0^+) \psi_L(0^+) \partial_x \psi_L(0^+)+h.c.\nonumber \\ &|\propto U_{2NR}| \cos (4\sqrt{\pi}\theta(0)+\alpha_{2NR}).\label{eq:2-fermion_NR}
\end{align}
Note that this term breaks $U(1)$ symmetry.

To study the full phase diagram we need to obtain RG equations near the NT fixed point. We only treat $U_{AR}$, $U_{NR}$, and $U_{2NR}$ in lowest order perturbation theory, but we are able to treat $U_{NT}$ exactly, using bosonization techniques. $U_{NR}$ similar to the previous section can be large and we can be starting at the NR fixed point at small other couplings.

We explore the phase diagram as a function of the Luttinger parameter $K$ and the bare value of $U_{NT}$. The phase diagram is shown for the $U(1)$ symmetric case in Fig. \ref{fig:regions_U1}, and for the $U(1)$ broken case in Fig. \ref{fig:regions}. The corresponding RG flows are shown in Figs. \ref{fig:RG_flow} and \ref{fig:3d_phase_diagram}.

When $U(1)$ symmetry is preserved, only one of the $t_1$ and $t_2$ terms can be present. 
As we showed above, then the Andreev reflection (AR) fixed point is stable for all $K<1$. For small $t$, the normal transmission (NT) fixed point is also stable in the green portion of the $U_{NT}-K$ phase diagram in Fig. \ref{fig:regions_U1}.
We argue below that this implies a non-trivial critical point, separating AR and NT fixed points, for $K<1/4$. 
We study this using ``$\epsilon$-expansion'' techniques when $K$ is slightly less than $1/4$.
 When $U(1)$ symmetry is broken the fixed point which is stable for all $1/4<K<1$ is the Andreev transmission (AT) one.
For $K<1/2$, there is another stable fixed point, normal reflection (NR), which results from $U_{NR}$ or $U_{2NR}$, with spontaneously broken time-reversal symmetry. For $K<1/4$ NR is the only stable fixed point. For $1/4<K<1/2$, the flows to these AT and NR points are separated by a surface in the $t_1$, $t_2$, $U_{NR}$ parameter space. Some details of these RG flows, related to the behavior in the $U(1)$ symmetric case, 
depend on $U_{NT}$.

\subsection{$U(1)$ symmetric case}

\subsubsection*{NT critical point}

We start the discussion of the $U(1)$ symmetric case by studying the weak coupling limit, corresponding to the NT fixed point. Let us obtain the weak-coupling RG near the NT fixed point. In this case it is important to take into account tunneling ($t_1$ for example), and the additional couplings allowed by both time-reversal and $U(1)$ symmetry, $U_{AR} (d^\dag d-1/2) \sin 2\sqrt{\pi}\phi(0)$ and $U_{NT}(d^\dag d - 1/2) \partial_x \phi(0)$. The RG equations are obtained by transforming the Hamiltonian using the unitary transformation $H'=\mathcal{U}^\dag H \mathcal{U}$, where 
\begin{align}\mathcal{U}=\exp[- i \sqrt{\pi}\tilde{U}_{NT} (d^\dag d - 1/2) \theta(0)].
\end{align} 
Here $\tilde{U}_{NT}=U_{NT}/uK$.
This transformation gets rid of the $U_{NT}$ term in the Hamiltonian, but introduces it into the tunneling terms:
\begin{align}
t_1 d^\dag \Gamma (e^{i \sqrt{\pi} (\phi-(1- \tilde{U}_{NT}) \theta)} + e^{i \sqrt{\pi}(-\phi - (1-\tilde{U}_{NT})\theta)}) + h.c., \\
t_2 d^\dag \Gamma (e^{i \sqrt{\pi}(\phi +(1+\tilde{U}_{NT}) \theta)} - e^{i \sqrt{\pi}(- \phi + (1+\tilde{U}_{NT}) \theta)}) + h.c.
\end{align}
The dimensions of $t_{1,2}$ ($\frac{1}{4}(K(1\pm \tilde{U}_{NT})^2 +1/K)$) and $U_{AR}$ ($1/K$) are easily deduced. We obtain the quadratic part of the RG equations in the appendix F. The result is:
\begin{align}
\frac{d \tilde{t}}{d \ell} &= \left[1 - \frac{1}{4}(K(1\mp\tilde{U}_{NT})^2 +1/K)\right]\tilde{t} + \tilde{t} \tilde{U}_{AR},
\nonumber \\
&&
\label{eq:U(1)t} \\
\frac{d \tilde{U}_{NT}}{d\ell} &= - 4 (\tilde{U}_{NT} \mp 1) \tilde{t}^2\label{eq:delta_F}\\
\frac{d \tilde{U}_{AR}}{d\ell} &= (1 - 1/K) \tilde{U}_{AR} + \frac{ \tilde{t}^2}{\pi} (1/K^2-(1\mp \tilde{U}_{NT})^2).\label{eq:AR}
\end{align}
for the properly normalized coupling constants ($\tilde{U}_{NT}=U_{NT}/uK$, $\tilde{t}= t_{1,2}\sqrt{2\pi}/\Lambda$, and 
$\tilde{U}_{AR} = U_{AR} 2 \pi/ \Lambda K$) and sign $-$ for $t_1$ and $+$ for $t_2$ cases correspondingly. Here $\ell = \ln \Lambda_0/\Lambda$, $\Lambda$ is the renomalized cutoff, and $\Lambda_0$ -- bare cutoff.

We see that for small initial $1-K$ and $\tilde{U}_{NT}$ the RG above signifies the relevance of the tunneling strength at the NT fixed point. Therefore the flow goes towards AR one. This is consistent with the simple analysis of the previous section.

\begin{figure}[t]
\includegraphics[width=0.7\linewidth]{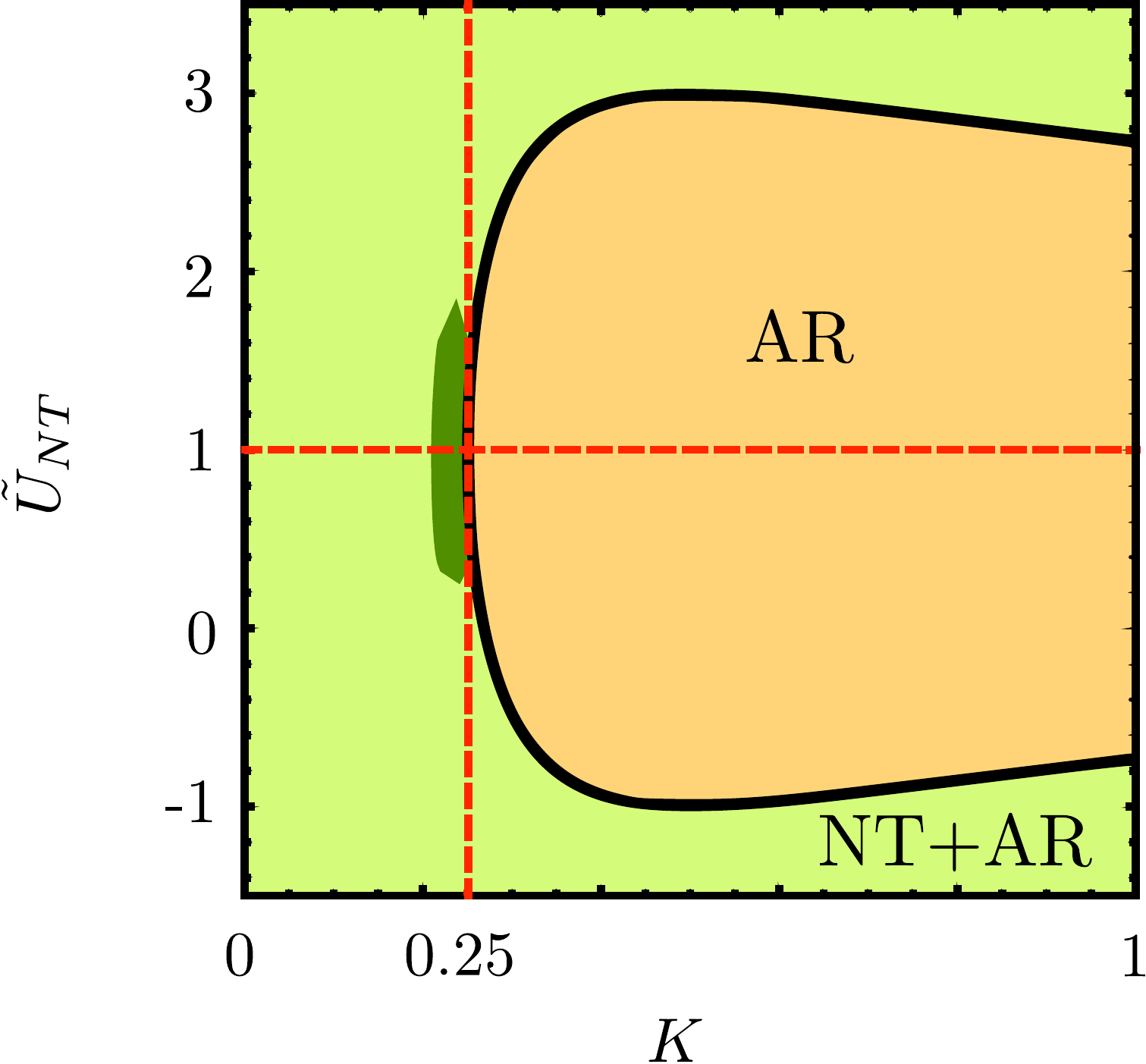}
\caption{Regions where the Andreev reflection (AR) critical point and both normal transmission (NT) and Andreev reflection (NT+AR) are stable in $\tilde{U}_{NT}$ and $K$ coordinates. We assume small $\tilde{U}_{AR}$ and $\tilde{t}$. $\epsilon$-expansion works in a small region of parameters near the line separating the two regions on the NT+AR side. The region is shown in darker color.
}\label{fig:regions_U1}
\end{figure}

 We now notice that when  $\tilde{t}$ is irrelevant, $\tilde{U}_{NT}$ does not renormalize much from its initial value due to \eqref{eq:delta_F}. Then the initial value of $\tilde{U}_{NT}$ controls the dimension of $\tilde{t}$: $ \frac{1}{4}(K(1-\tilde{U}_{NT})^2 +1/K)$.
Since, as we have shown above, the AR fixed point is stable for all $K<1$, we expect a critical surface to exist in the 
$(t,U_{NR}, U_{NT})$ space, separating flows to these competing stable fixed point. The nature of this critical surface is qualitatively different for $K<1/4$ and $K>1/4$.

\begin{figure}[t]
\includegraphics[width=0.8\linewidth]{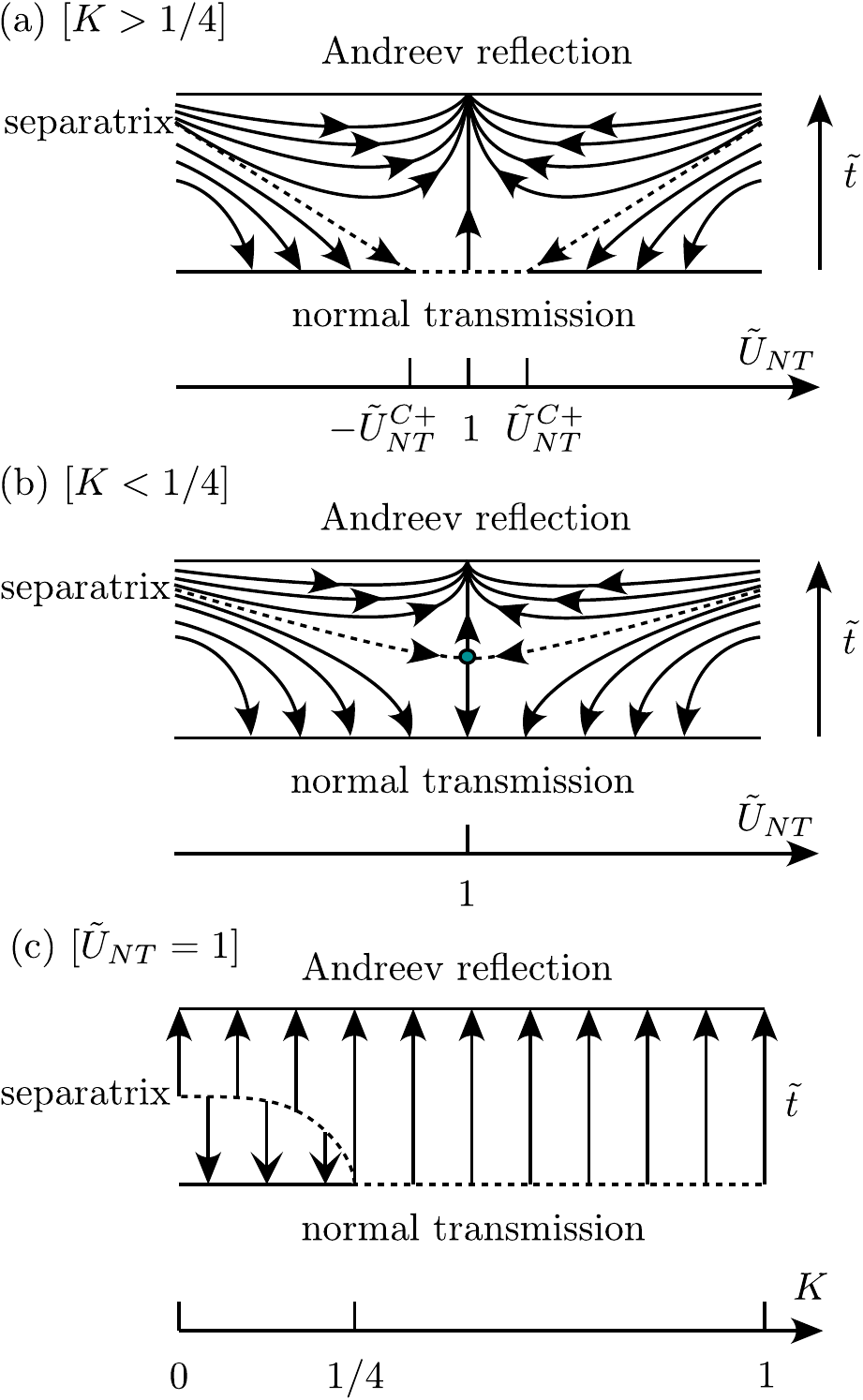}
\caption{Renormalization group flow for the $U(1)$ symmetric case, $K<1$. (a): RG flow in the $\tilde{U}_{NT}$-$\tilde{t}$ space for $K>1/4$. $\tilde{U}_{NT}^{C\pm}$ represent the boundaries; outside the interval between them there are two stable fixed points, NT and AR (see Fig. \ref{fig:regions_U1}). Flowing to $\tilde{U}_{NT}^{C\pm}$ is the separatrix dividing the flow to the NT and AR fixed points. For $K>1/4$ there is no non-trivial fixed points, only NT and AR. (b): same as (a), but now $K<1/4$. There are always two stable critical points, NT and AR. There is a separatrix line flowing to the non-trivial fixed point, corresponding to non-zero values of $\tilde{t}$ and $\tilde{U}_{AR}$, see the main text for details. (c): RG flow for $\tilde{U}_{NT} = 1$ with the line of non-trivial fixed points separating AR and NT fixed points starting at $K=1/4$. In this case the separatrix consists of non-trivial fixed points for each $K<1/4$ from (b).
}\label{fig:RG_flow}
\end{figure}

For all $K>1/4$ and large enough $\tilde{U}_{NT}$ (see fig. \ref{fig:regions_U1}) there is a critical surface separating the flows to the AR and NT fixed points. It is clear that the NT fixed points are at $\tilde U_{NT}\neq 1$, since for $\tilde{U}_{NT}=1$ $\tilde{t}$ is a relevant perturbation and the system flows to the AR fixed point. However, for $|\tilde U_{NT}-1|$ sufficiently large, $\tilde t=0$ is a stable fixed point. Therefore, we conclude that the RG flow on the critical plane is to the NT fixed point with 
$t=0$, $U_{AR}=0$ and $\tilde U_{NT}$ given by the solutions of
\be {1\over 4}\left[ K(1-\tilde U_{NT})^2+{1\over K}\right]=1,\ee
which is
\be \tilde U_{NT}^{C\pm}=1\pm {1\over K}\sqrt{4K-1}.\ee
 The RG flow in that case is schematically shown in Fig. \ref{fig:RG_flow}a.

For all $K<1/4$ the RG flow on the critical surface is to a non-trivial critical point where $\tilde U_{NT}=1$, and 
$t$ and $U_{AR}$ are non-zero. 
For $K = 1/(4 + 4\epsilon)$, $0<\epsilon\ll 1$, we may find this critical point from the lowest order RG equations of 
Eq. (\ref{eq:U(1)t})-(\ref{eq:AR}), which may be written:
\bea 
\frac{d\tilde{t}}{d\ell} &=& - \epsilon \tilde{t} + \tilde{U}_{AR} \tilde{t},\label{eq:critical_t} \\
\frac{d\tilde{U}_{AR}}{d\ell} &=& - (3 + 4\epsilon) \tilde{U}_{AR} + \alpha \tilde{t}^2.\label{eq:critical_U}
\eea
Here we denote $\alpha = \frac{1}{\pi K^2}$.

Before proceeding let us show that we have not missed the important terms in the equation. For that we notice that for small $\epsilon$: $\tilde{U}_{AR}\sim\epsilon$ and $\tilde{t}\sim\sqrt{\epsilon}$. Additionally, let us mention the irrelevant term we did not take into account so far, $\propto t_3 d^\dag e^{i\sqrt{\pi}(3\phi + (1 - \tilde{U}_{NT})\theta)}$. The RG equation for $t_3$ cannot have $\tilde{t}^2$ term in it, as then the powers of $d$ do not add up. Therefore $t_3\lesssim \epsilon^{3/2}$ and it cannot change the position of the non-trivial fixed point up to the order of $\epsilon$. There is an additional term that might appear missing from the equations, $\tilde{t}^3$. This is because on the line $\tilde{U}_{NT}=1$ where the critical point is situated $t\propto i(d-d^\dag) (e^{i\sqrt{\pi}\phi} + e^{-i\sqrt{\pi}\phi})$. This corresponds to the case of single Majorana coupled to a single channel in \onlinecite{Aff13}, where no third order term in the RG equation is obtained. Therefore we have taken into account all the terms of the order of $\epsilon^{3/2}$.

But let us return to the RG equations near the barely stable normal transmission fixed point, \eqref{eq:critical_t} and \eqref{eq:critical_U}. The equations support an unstable critical point $\tilde{U}_{AR}=\epsilon$, $\tilde{t} = \sqrt{(3+4\epsilon) \epsilon/\alpha}$, which starts off being close to the normal transmission fixed point, but then branches off. The RG flow for $K<1/4$ is shown in the 
$(\tilde t,\tilde U_{NT})$ plane in Fig. \ref{fig:RG_flow}b.
The dotted line in Fig. \ref{fig:RG_flow}c is a line of non-trivial critical points, separating AR and NT. We present the detailed discussion of the RG flow near the non-trivial critical point in the Appendix E, from which we obtain the flow towards the non-trivial critical point along the separatrix surface defined by the equation:
\begin{align}
\tilde{t} = \sqrt{\frac{3\epsilon}{\alpha} + \frac{3(1-\tilde{U}_{NT})^2}{208} - \frac{2\epsilon(\tilde{U}_{AR}-\epsilon)}{\alpha}}.\label{eq:tildet}
\end{align}

\subsection{Interacting $U(1)$ breaking case}

\begin{figure}[t]
\includegraphics[width=0.7\linewidth]{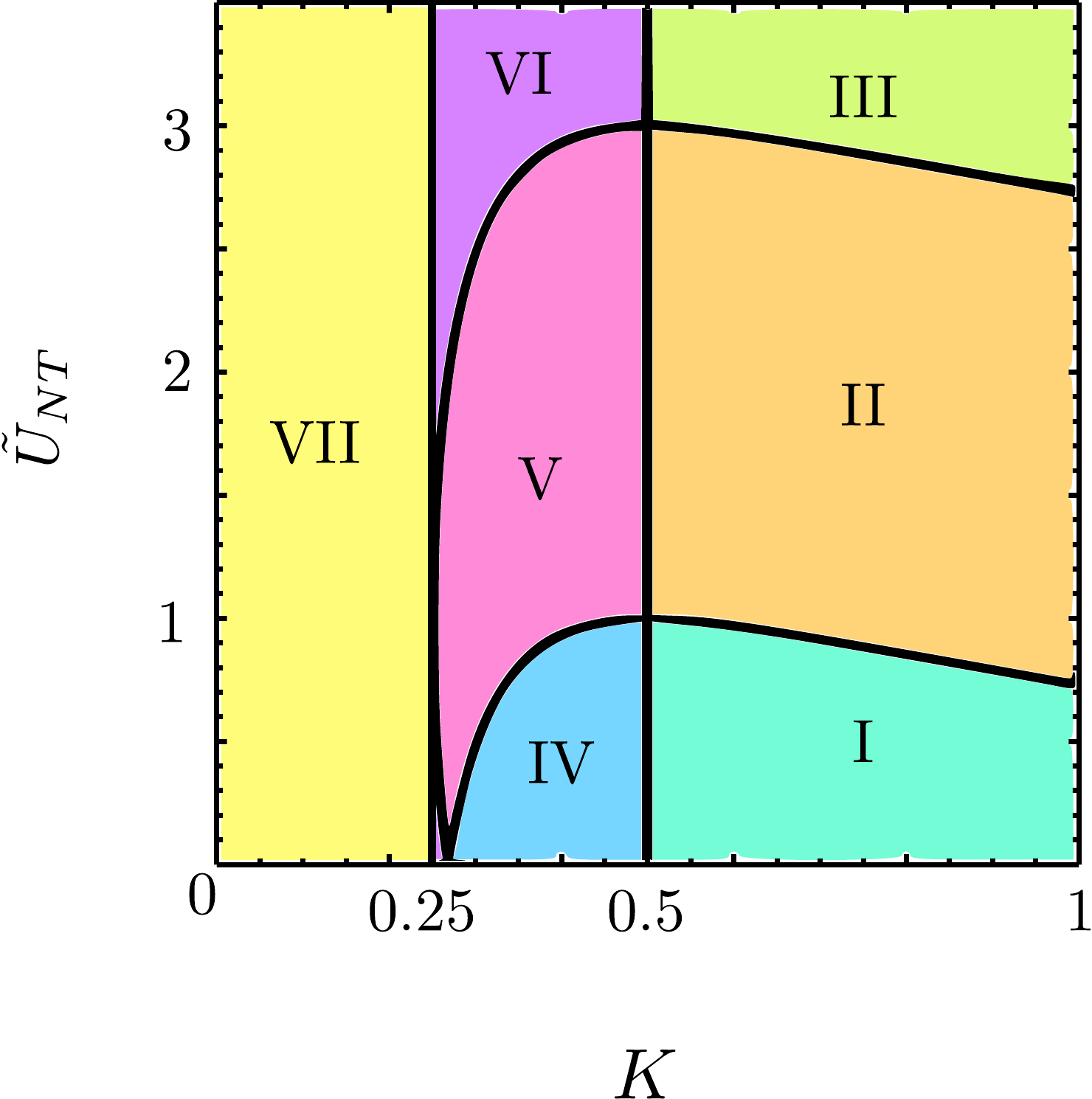}
\caption{Seven regions of the phase diagram corresponding to the different RG flow in Fig. \ref{fig:3d_phase_diagram}. They are separated by lines: $K=1/2$, where the normal reflection (NR) critical point becomes stable; $K=1/4$, where the Andreev transmission (AT) fixed point becomes unstable; $\frac{1}{4K}+ \frac{K}{4}(1+\tilde{U}_{NT})^2=1$ (top line), where $t_2$ becomes irrelevant; $\frac{1}{4K}+ \frac{K}{4}(1-\tilde{U}_{NT})^2=1$ (bottom line), where $t_1$ also becomes irrelevant. The phase diagram for $\tilde{U}_{NT}<0$ is obtained from above by replacing $t_1\leftrightarrow t_2$ and $\tilde{U}_{NT} \to - \tilde{U}_{NT}$. This figure can be though of as two overlayed Fig. \ref{fig:regions_U1} with $\tilde{U}_{NT}$ for $\tilde{t}_1$ and $-\tilde{U}_{NT}$ for $\tilde{t}_2$, with additional $K=1/2$ and $K=1/4$ lines separating the stability regions of the NR and AT fixed points correspondingly.
}\label{fig:regions}
\end{figure}
\begin{figure*}
\includegraphics[width=0.95\linewidth]{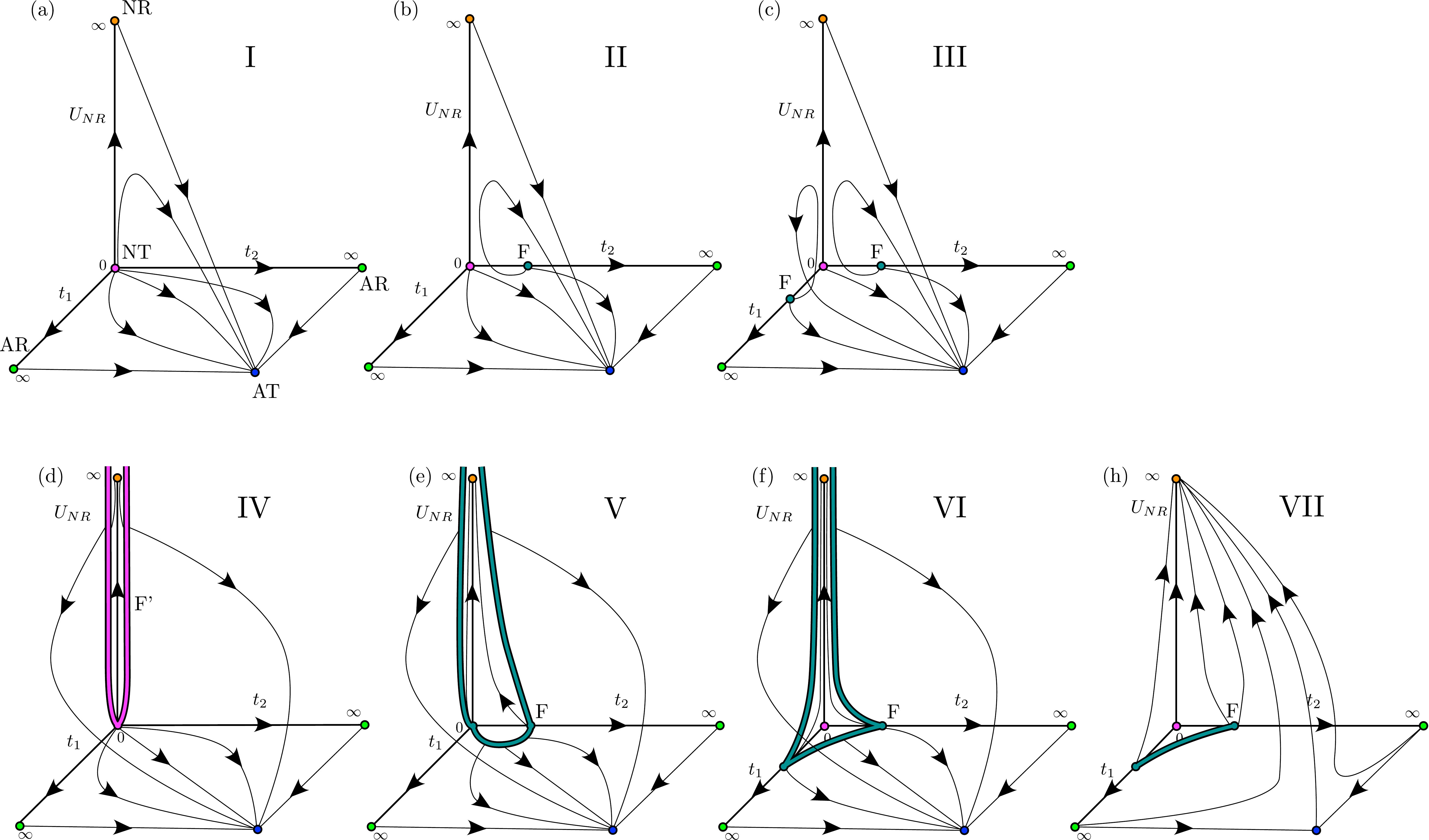}
\caption{Hypothesized phase diagrams of the model for different values of $K$ and $\tilde{U}_{NT}$, corresponding to different regions of Fig. \ref{fig:regions}. The flow is based on the weak-coupling RG near the trivial critical points. (a): region I in Fig. \ref{fig:regions}, where the only stable fixed point is AT fixed point. (b): II, where the NT fixed point becomes stable for $U(1)$ symmetric case, where only $t_2$ is present. (c): III, where the NT fixed point is stable for both $U(1)$ symmetric cases. (d): IV, where NR fixed point becomes stable, but both hoppings are relevant and NT fixed point and the flow along the $U(1)$ symmetric line goes towards AR fixed point. In this case we can also show that the flow along the plane separating the AT and NR fixed points is towards strong coupling (e): V, where NR fixed point is stable, and at the same time in the $U(1)$ symmetric case when only $t_2$ is present, NT becomes stable. (f): VI, where NR is stable and in both $t_1$ and $t_2$ $U(1)$ symmetric directions the flow is towards NT fixed point. (h): VII, where NR is the only stable fixed point, however the unstable fixed points on the $U(1)$ line are still present. We hypothesize that they are connected by the line of unstable critical points, similar to region VI.
}\label{fig:3d_phase_diagram}
\end{figure*}

\subsubsection{Normal reflection fixed point}

In the $U(1)$ symmetry breaking case the additional $U_{NR}$ and $U_{2NR}$ terms are allowed, see eqs. \eqref{eq:Jz} and \eqref{eq:2-fermion_NR}. $U_{NR}$ has scaling dimension $K$ at the NT fixed point and is relevant for all $K<1$. $U_{2NR}$ has the dimension $4K$ and is always less relevant than $U_{NR}$ at the NT critical point. However, at AT fixed point $U_{NR}$ is not present since $d$ level is screened. $U_{2NR}$ is allowed however, and it becomes relevant for $K<1/4$, making AT critical point unstable.

We have discussed the case of $U_{NR}$ growing to infinity in the previous section. Here we notice that
$U_{2NR}$ can be present at this fixed point but it may only change the value of pinning $\theta$.

Consider now the case when the coefficient of $U_{2NR}$ grows to infinity. The term $U_{2NR}\cos (4\sqrt{\pi}\theta+\alpha_{2NR})$ pins $\theta$ at:
\begin{align}
\theta + \frac{\alpha_{2NR}}{4\sqrt{\pi}} &= -3\sqrt{\pi}/4, -\sqrt{\pi}/4, \sqrt{\pi}/4, 3\sqrt{\pi}/4.
\end{align}
At first glance the fixed point is different from the previously discussed in this section, since it does not involve $d$ operator at all. However both the marginal operator $\partial_x\theta$ and $U_{NR}$ are also allowed to be generated in RG. First tunes through the values of pinning of $\theta$, and the other combined with the pinned $\theta$ fixes the occupation of $d$ level. Therefore, there is only one line of fixed points, connecting the different pinnings of $\theta$, along which occupation of $d$ can be either $0$ or $1$.

\subsubsection{Normal transmission critical point, non-trivial critical points and discussion of global phase diagram}

Finally we turn to the analysis of stability of the NT fixed point. As we do not have $U(1)$ symmetry anymore, the term \eqref{eq:Jz} is allowed and is always relevant at the NT critical point. (in $U(1)$-symmetric case it is forbidden and the NT fixed point may become stable). 

Even if $U_{NR}$ is absent from the initial Hamiltonian, it gets generated under RG except for $U(1)$ symmetric lines. This means that in the general $U(1)$-breaking case the stability of the NR fixed point, discussed in the previous section, influences the global phase diagram. Notice that the equations for $\tilde{t}_1$ and $\tilde{t}_2$ are related by the time-reversal symmetry in the leads, which leaves  $d$ unchanged, i.e.:
\begin{align}
d &\to d, \\
\psi_L &\to - i \psi_R, \\
\psi_R &\to i \psi_L.
\end{align}
This sends $t_1$ to $t_2$ and $\tilde{U}_{NT}$ to $-\tilde{U}_{NT}$. 

So, with broken $U(1)$ symmetry, NT is never stable.  For $K>1/2$ the only stable fixed point is AT, corresponding 
to cases I, II and III in Fig. \ref{fig:regions}.  For $1/4<K<1/2$ 
NR is also stable so there must be a critical surface separating these two stable fixed points, corresponding to cases 
IV, V and VI in Fig. \ref{fig:regions}.  Depending on the value of $\tilde{U}_{NT}$ and $K$ one or both of the 
critical points on the $t_1$ and $t_2$ axes may be stable in the U(1) symmetric case when the other $t_i$ 
and also $U_{NR}$ are zero. Finally, for $K<1/4$ the AT critical point becomes unstable, and globally the RG flow leads to the NR line of fixed points. However, there are still critical points on the $U(1)$ symmetric lines, and presumably a line connecting them as the impurity entropy is the same at the two points. Additionally, the NR operator is irrelevant at AR critical point, therefore the flow first goes towards AT critical point and only then turns towards NR, as shown in the panel VII of Fig. \ref{fig:3d_phase_diagram}. This leads to the 7 different cases shown in Fig. \ref{fig:regions} and 
\ref{fig:3d_phase_diagram}.

\section{Conductance\label{sec:conductance}}
Having obtained the predictions for the scattering matrix in the non-interacting case and the fixed points in the interacting cases, we will now describe the physically observable quantity, conductance. We consider our device to be a 3-lead junction.  Two of the leads, labeled $1$ and $2$ in Fig. \ref{fig:setup}a) and \ref{fig:setup}b), correspond to two 
ends of the QSH edge where contacts are applied and the third, labeled $S$,  to the superconductor. We consider infinitesimal voltages, $V_i$ applied to each of these leads and 
measure the current, $I_j$ flowing towards the junction in lead $j$. The (linear) conductance tensor is defined by 
\be I_i=\sum_{j=1}^3G_{ij}V_j.\ee
Charge conservation implies that the total current flowing into the junction must be zero so 
\be \sum_i G_{ij}=0.\ee
Requiring that the currents in each wire vanish when all 3 voltages are equal implies
\be \sum_j G_{ij}=0.\ee
These 2 conditions imply that $G_{Sj}$ and $G_{jS}$, for $j=1$, $2$, are determined by $G_{ij}$ for $i,j\in \{1,2\}$. 
Furthermore it follows from the time-reversal symmetry that $G_{11}=G_{22}$ and $G_{12}=G_{21}$. Therefore we only need to calculate 2 independent 
components of the conductance tensor, $G_{11}$ and $G_{12}$. 

\subsection{Non-interacting case}
In this case we can calculate the conductance from the S-matrix calculated in Sec. II, using a generalisation of the Blonder-Tinkham-Klapwijk (BTK) formula.\cite{Blonder82}
A parameterization of this $S$-matrix in terms of  electrons /holes and left/right movers was given in Eq. (\ref{SMlab}).  It is convenient to relabel the  
$S$-matrix elements according to which lead, $1$ or $2$ the electron/hole terminates/originates from. 
This corresponds to
\bea
S_{11}&=&S_{LR}, \qquad S_{12}=S_{LL},\nonumber \\ 
S_{22}&=&S_{RL}, \qquad S_{21}=S_{RR}.
\eea
Then, following BTK,  the conductance in linear response can be written as:
\begin{align}
G_{ij}&=\int{d\eps}[-f'(\eps)]g_{ij}(\eps), \\ 
g_{ij}(\eps)&=\delta_{ij}-\abs{S_{ij}^{ee}(\eps)}^2+\abs{S_{ij}^{he}(\eps)}^2,\label{gij}
\end{align}
where $f'(\epsilon)$ is the derivative of the Fermi function. and we work in units where $e^2/h=1$.  Note that $S^{ee}_{LR}=S^{ee}_{11}=0$, corresponding to zero normal reflection 
amplitude due to time-reversal. Thus $G_{11}$, giving the current flowing from contact $1$ due to a voltage applied at contact $1$, is a sum 
of two positive terms: one corresponding to the electron emitted from the contact and the other to the Andreev reflected hole. 
$G_{21}$, the current flowing from contact $2$ due to a voltage applied at contact $1$ also has two contributions: a negative  
one due to normal transmission of an electron and a positive one corresponding to Andreev transmission. 

\subsubsection{$T=0$}
For the zero energy conductance one can obtain very simple expressions using the applicability of \eqref{eq:S} for arbitrary $U_1$ and $U_2$:
\begin{align}
G_{11} = G_{22} = 1 + \left(\frac{t_1^2 - t_2^2}{t_1^2 + t_2^2}\right)^2,\\
G_{12} = G_{21} = \frac{4 t_1^2 t_2^2}{(t_1^2 + t_2^2)^2}.
\end{align}
For the $U(1)$ symmetric case, where $t_1$ or $t_2=0$, $G_{21}=0$ since there is pure Andreev reflection so that a voltage applied to contact $1$ leads 
to no current flowing from contact $2$. When $t_1=t_2$, $G_{11}=G_{21}=1$, corresponding to perfect Andreev transmission.  A voltage 
applied to lead 1 leads to only a particle current in lead 1 and only a hole current in lead 2. 
 Another property worth noting is that $G_{11} + G_{21}=2$ for any $t_1$ and $t_2$ at zero temperature.
  This is true, since only Andreev processes happen at zero energy -- Andreev reflection and Andreev transmission.
  Note that the limiting cases coincide with the results in Fig. \ref{fig:numerics}.

\subsubsection{$U_i=0$}
For $U_i=0$ one can also write simple conductance expressions following from \eqref{SU0}:
\begin{align}
g_{11} =  1 + |(z-1)/2|^2\left(\frac{t_1^2 - t_2^2}{t_1^2 + t_2^2}\right)^2,\\
g_{21} = |(z-1)/2|^2 \frac{4 t_1^2 t_2^2}{(t_1^2 + t_2^2)^2} - |(z+1)/2|^2.
\end{align}
Here $z\equiv \frac{i v_F - 2 t^2/(v_F k)}{i v_F + 2 t^2/(v_F k)}$. Notice that for $t_1=t_2$,  $g_{ 11}$ does does not depend on energy, implying that $G_{11}$ 
is temperature-independent. This is a consequence of the fact that only Andreev reflection affects  $g_{11}$,  as we see from Eq. (\ref{gij}), 
and the fact that no Andreev reflection occurs at any energy in this case, 
an exact consequence of the non-local $U(1)$ symmetry that appears in the non-interacting model in this case, 
discussed in Section IIB).

\subsubsection{Numerical solution}
\begin{figure}[]
\includegraphics[width=\linewidth]{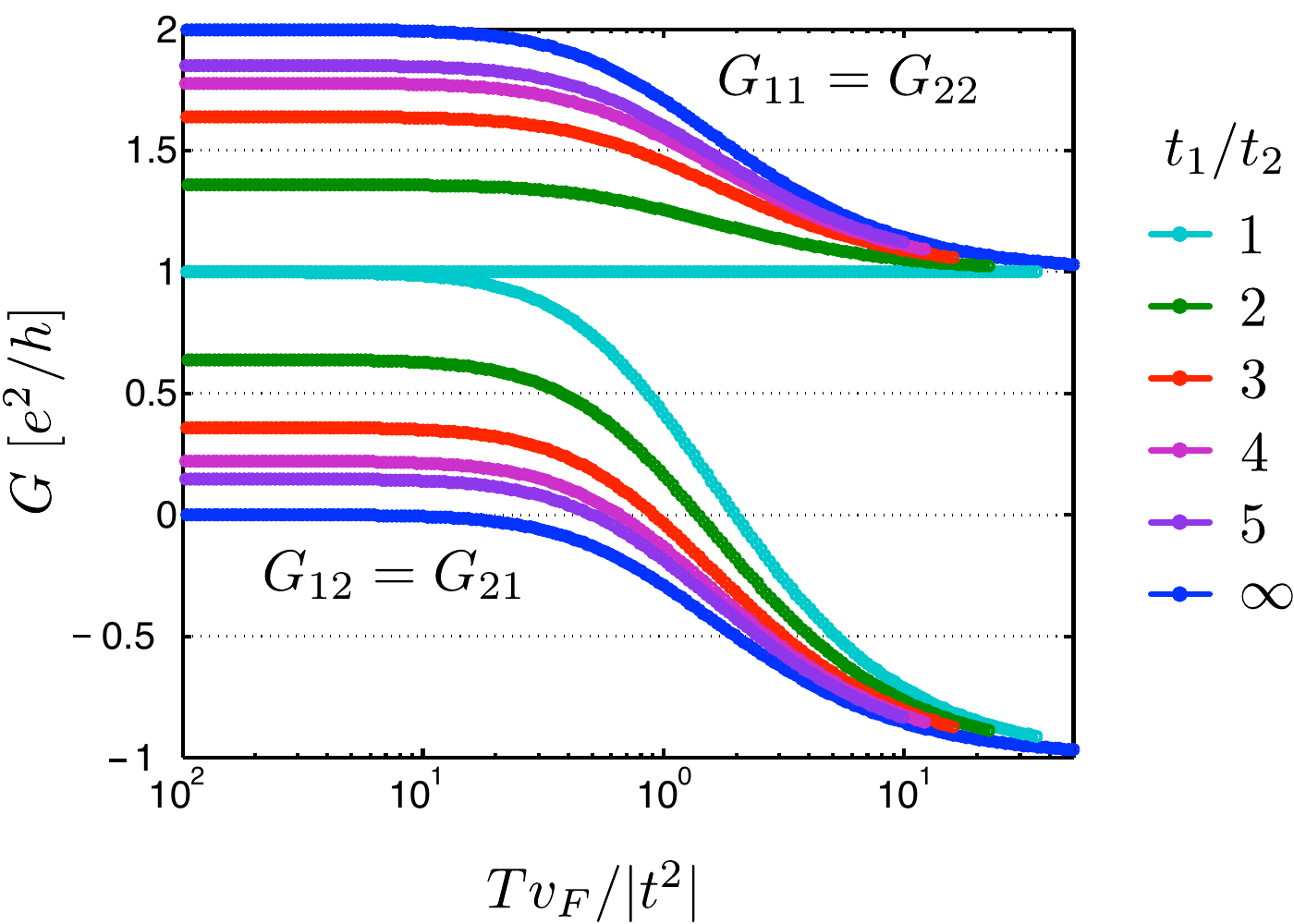}
\caption{Conductance tensor as a function of temperature for the non-interacting case. Different colors correspond to different ratios $t_1/t_2$, simulation is performed for $U_1=U_2=0$. These ratios run from $1$, corresponding to full Andreev transmission to $\infty$, corresponding to full Andreev reflection.}\label{fig:numerics}
\end{figure}

More generally, at the IR fixed point $g_{11}$  can take any value between $0$ and $1$ and $g_{12}$ can take any value between $-1$ and $1$.
Numerical results for $U_1=U_2=0$ and different values of $t_1/t_2$ are shown in Fig. \ref{fig:numerics}.

\subsection{Corrections to the conductance due to trivial conductance paths}

\begin{figure}[t]
\includegraphics[width = \linewidth]{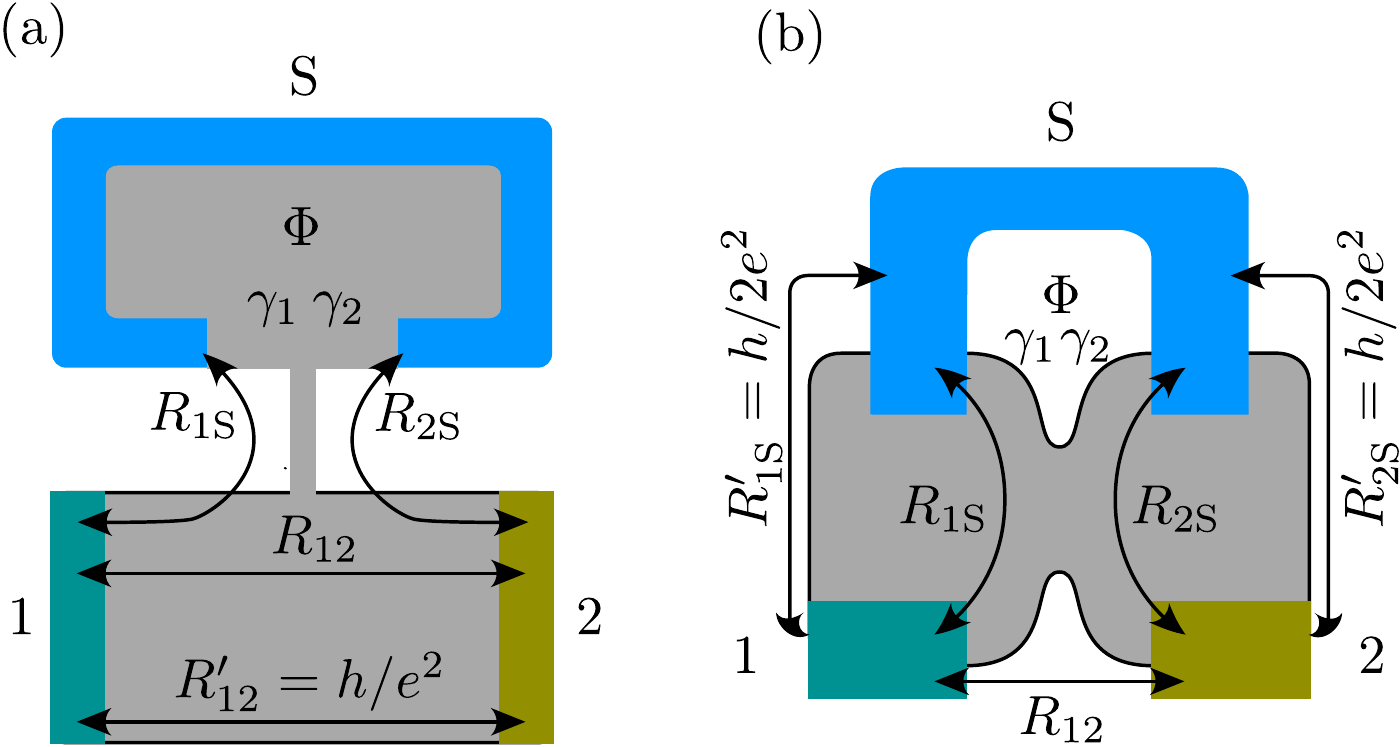}
\caption{Resistance networks for the two setups with trivial conductance paths.
}\label{fig:schemes}
\end{figure}

We replot the setups shown in Fig. \ref{fig:setup} in a schematic way showing different current paths between the leads, 1, 2, and SC. Each connection is replaced by a resistor. This is possible since we consider the leads to be macroscopic and in good contact with the edge states. This gets rid of any possible phase coherence under the leads and different current paths can be considered separately. 

We know the conductances along the Majorana paths in Fig. \ref{fig:schemes}; those can be easily transformed into resistances shown there. However, in parallel with the interesting current paths there are inevitable additional paths due to the topological nature of the QSHE and presence of the edge channels on all the edges of the device. These are $R_{\rm 12}'$ in Fig. \ref{fig:schemes}a, and $R_{\rm 1S}'$, $R_{\rm 2S}'$ in Fig. \ref{fig:schemes}b. 

Treating these trivial conductance paths experimentally can be done in two different ways.
 If one keeps them in the quantized regime, these conductance paths have known resistances: $R_{\rm 12}'=h/e^2$, as it is a single normal channel; $R_{\rm 1S}'=R_{\rm 2S}'=h/2e^2$, as it is a single path with perfect Andreev reflection from the superconductor. Therefore whatever is the measured resistance between the leads, one can subtract their contribution as described above. 

Another way to treat the problem is by closing down the unwanted paths. This is possible for example by applying local magnetic field in the quantum dot setup discussed in [\onlinecite{SMi13}] or by making the corresponding trajectories much longer than the mean free path due to inelastic processes in the QSH edge. In realistic systems it is of order of few $\mu m$.\cite{Kon07, LDu15} The second method has the advantage of explicitly preserving time-reversal symmetry.

\subsection{Interacting case}

\begin{figure*}[t]
\includegraphics[width=\linewidth]{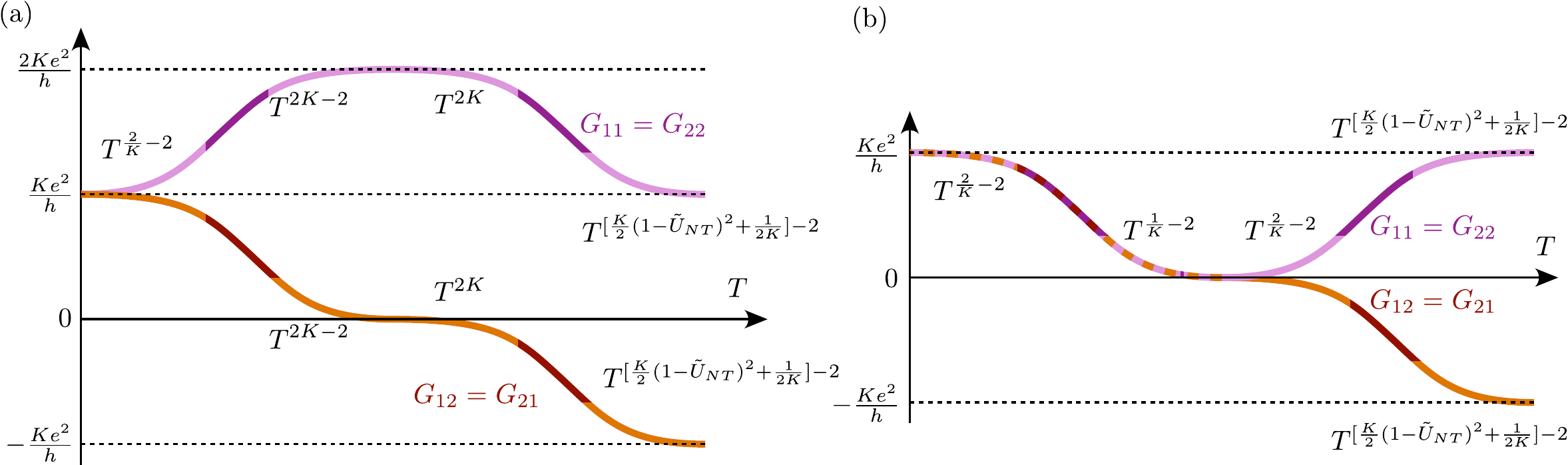}
\caption{Conductance as a function of temperature, purple line shows $G_{11}=G_{22}$, and orange line -- $G_{12}=G_{21}$. 
We assume $K<1$. (a) Almost $U(1)$ symmetric case. We assume that the system starts at the normal transmission (NT) fixed point, and the tunneling is relevant at NT fixed point, $\frac{K}{4}(1-\tilde{U}_{NT})^2 + \frac{1}{4K}<1$. Starting from NT the system flows to AR fixed point. The AR fixed point is unstable and the system flows to AT fixed point after that .(b) $U(1)$-breaking case but NR is unstable. We also assume that the $\tilde{U}_{NR}$ is the most relevant operator at the NT fixed point, i.e. $(1-\tilde{U}_{NT})^2 + \frac{1}{4K}>K$, but NR fixed point is unstable, i.e. $K>1/2$. Then the system first flows to NR fixed point and only then turns to the stable AT fixed point. The dependence of the conductance on temperature near the fixed points is given by the dimensions of corresponding most relevant or irrelevant operators, see eq. \eqref{eq:G_dimension}.
}\label{fig:conductance}
\end{figure*}

The case of  interacting electrons requires a more careful treatment to obtain the conductance. In this sub-section we consider the case of  interacting Luttinger liquid leads, but 
we discuss the case of Fermi liquid leads in sub-section VD. We  first consider the conductance tensor at the 3 fixed points discussed in Sec. II: normal transmission, Andreev reflection 
and Andreev transmission. 

The normal transmission fixed point corresponds to $t_i=U_i=0$ and a perfectly translationally invariant wire. Following the bosonization conventions introduced 
in sub-section IIIA, the current operator at the fixed points, where the Hamiltonian is quadratic in bosons, is:
\be J(x)=-K{v_F e\over \sqrt{\pi}}\partial_x\theta (x).\ee
We obtain the conductance from the Kubo formula, where a voltage is applied at a point $x$ and the current 
is measured at a point $y$:
\be G=\lim_{\omega \to 0}{1\over \omega}\int_0^\infty dt e^{(i\omega -\delta )t}\langle J(x,t),J(y,0)]\rangle\label{Kubo}\ee
Here $J(x, t)$ is the current operator at position $x$ and time $t$.
The retarded Green's function for the current operator is:
\be G_{\hbox{ret}}(x-y,\omega )={Kv_F^3\over \pi}\int_{-\infty}^\infty {dk\over 2\pi}{k^2e^{ik(x-y)}\over (\omega +i\delta )^2-v_F^2k^2}\ee
yielding
\be G=K.\ee
Because of our sign convention, with all  currents directed towards the junction, this corresponds to
\be G_{11}=-G_{12}=K.\ee

The Andreev transmission fixed point  corresponds to the boundary condition of Eq. (\ref{bcFAT}). To see how this changes the conductance from 
the NT case note that it is equivalent to normal transmission if we redefine the fermion fields at $x>0$ by:
\be \psi_{L/R}(x)\to \psi_{L/R}^\dagger (x).\ee
This has the effect of changing the sign of the current operator at $x>0$:
\be J=\psi^\dagger_R\psi_R-\psi^\dagger_L\psi_L\to \psi_R\psi^\dagger_R-\psi_L\psi^\dagger_L=-(\psi^\dagger_R\psi_R-\psi^\dagger_L\psi_L).\ee
This change in sign of the current at $x>0$ corresponds to a change of sign of $G_{12}$
\be G_{11}=G_{12}=K.\ee

The Andreev reflection fixed point corresponds to the boundary condition
\be \psi_L(0^\pm )=-\psi_R^\dagger (0^\pm ).\ee
In this case clearly $G_{12}=0$ since a voltage applied to contact $1$ does not lead to any current flowing from contact $2$. To calculate the 
current  flowing from contact $1$ due to a voltage applied at contact $1$ we can use the AR boundary condition to unfold the right movers in the $x<0$ half of the system, defining
\be
\psi_L(x)\equiv-\psi_R^\dagger (-x),\ \  (x<0).\ee
This has the effect, for $x<0$:
\bea J(x)&\to& :\psi_R^\dagger (x)\psi_R(x):-:\psi_R(-x)\psi_R^\dagger (-x):\nonumber \\
&=&J_R(x)+J_R(-x) \label{JAR}
\eea
where $J_R$ is the current carried by right movers.  The term in Eq. (\ref{Kubo}) coming from the right-moving term in the current is
\be G_{\hbox{ret},R}(x,y)=K\Theta (x-y).\ee
Thus Eq. (\ref{JAR}) implies 
\be G_{11}=K[\Theta (x-y)+\Theta (y-x)+\Theta (x+y)+\Theta (-x-y)]=2K.\ee
In all three cases, the conductance tensor simply gets multiplied by a factor of $K$ compared to the non-interacting case. 

In the interacting case the fourth possibility arises, fixed point with spontaneously broken time-reversal symmetry, NR. It corresponds to absence of transport between the leads and into the superconductor. Therefore the conductance tensor is identically zero at the NR fixed point.

We expect the temperature dependence of the conductance near these 4 simple fixed points to be controlled by the leading relevant or irrelevant operators. 
In each case, these make a contribution to the conductance in 2nd order perturbation theory when the corresponding coupling constants are small, 
corresponding to being near the fixed point. We can estimate the asymptotic temperature dependence in each case from the dimension of the operator. 
An operator of RG scaling dimension $d$ contributes a term to the conductance in 2nd order perturbation theory scaling as:
\be \delta G\propto T^{2(d-1)}.\label{eq:G_dimension}\ee
For irrelevant operators, $d>1$, this correction vanishes as $T\to 0$. However, for relevant operators, $d<1$ this correction blows up as the 
temperature is lowered, signifying the RG flow away from the fixed point. 

For general small values of the $t_i$'s, $\tilde{U}_{NT}\approx 1$, and $K>1/2$, (region I of the phase diagram in Fig. \ref{fig:regions}), we expect a simple flow from the Normal Transmission fixed point to the Andreev Transmission fixed point. If we are in regions II and III of phase diagram, we expect first flow towards NT fixed point, and only then eventually to AT fixed point. If we are in IV-VI regions and have small $t_i$'s, we flow to normal reflection fixed point.  Most of the dimensions needed to predict 
the exponents of the flow with temperature were worked out in Sec. III. The two new ones we need are the leading irrelevant operators at NR and AR fixed points. 

We start from the AR fixed point, where we consider the leading irrelevant operator in the presence 
of $U(1)$ symmetry.  We look for the lowest dimension $U(1)$ and time-reversal invariant operator which couples $x<0$ to $x>0$ at the AR fixed point, using 
the $C_L$ transformed operators $\tilde  \psi_{L/R}(0^\pm )$ and imposing the corresponding $AR$ boundary condition, which corresponds 
to normal reflection in this basis, Eq. (\ref{bc}). While the operator $\tilde \psi_R^\dagger (0^+)\tilde \psi_R(0^-)$ cannot occur due to time-reversal symmetry, as 
discussed in Sec. IIB), 
an operator of the form  
\be i[\tilde  \psi_R^\dagger (a)+\tilde  \psi_L^\dagger (a)][\tilde  \psi_R(-a)-\tilde  \psi_L(-a)]+h.c.,\ee
where $a$ is a short distance scale of the order the lattice constant, is invariant under time-reversal which takes the form
given in Eq. (\ref{TRGB})  in this basis. Imposing the boundary condition, and Taylor expanding this becomes
\be\propto  i\partial_x\tilde  \psi_R^\dagger (0^+)\psi_R(0^-)\ee
of dimension $(1+K)$.

We continue with considering the leading irrelevant operator at the NR fixed point. We notice that the flow exactly to the fixed point happens when $t_i=0$, or the number conservation is present. Then the situation is the same as the well-known problem of single potential scatterer\cite{Kane92}, where the leading irrelevant operator is the tunneling between the two sides of the quantum wire, separated by the impurity. The tunneling has the form:
\begin{align}
\psi^\dag_R(-0)\psi_R(+0) \propto e^{i \sqrt{\pi} (\phi(+0) - \phi(-0))},
\end{align}
where we have used that $\theta$ is fixed at NR fixed point. This operator has the dimensions $1/K$.

In Fig. \ref{fig:conductance} we show the scaling behavior of conductance starting from weak coupling (NT) fixed point. In Fig. \ref{fig:conductance}a we show $U(1)$ preserving case, where the RG flow is between the NT and AR fixed points for in the region where the AR is the only stable fixed point, see Fig. \ref{fig:regions_U1}. In Fig. \ref{fig:conductance}b we show the $U(1)$ breaking case, corresponding to the region III of \ref{fig:regions}, where AT is the only stable fixed point, but tunneling is irrelevant for small $t$. The flow between the fixed point goes as NT$\to$NR$\to$AT, and we depict the corresponding scaling behavior of conductance.

\subsection{Fermi liquid leads}

\begin{figure}[t]
\includegraphics[width = 0.6\linewidth]{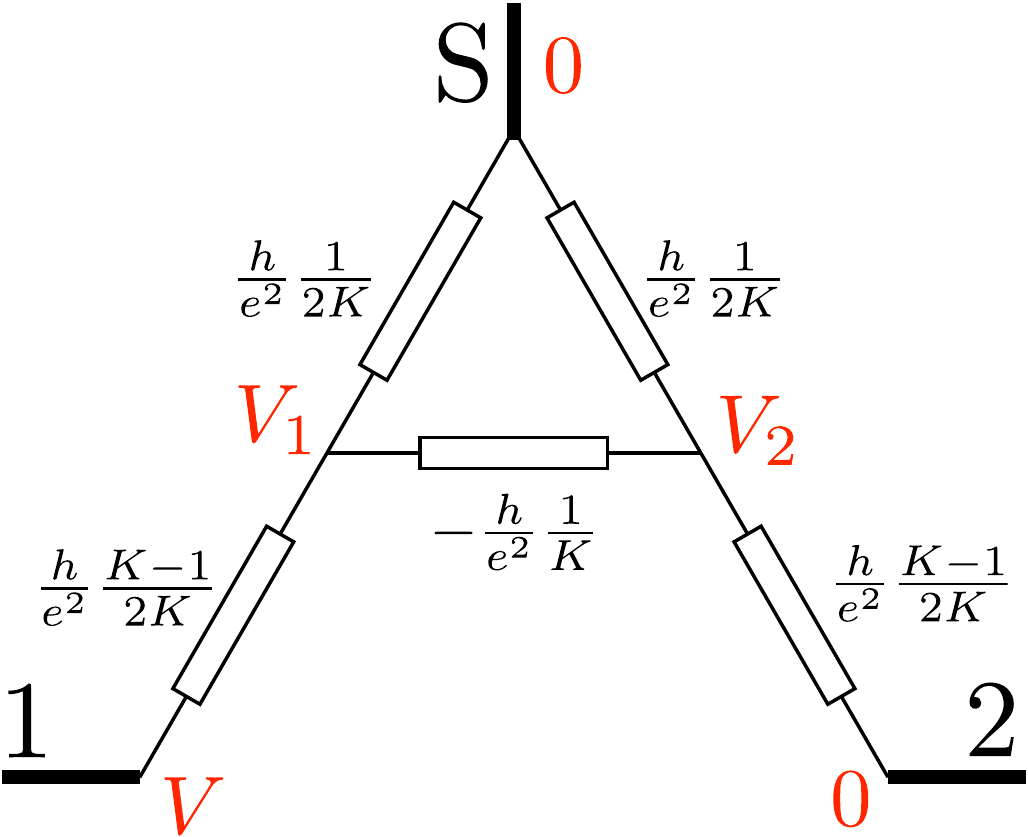}
\caption{Resistance network which model the system in presence of the Fermi liquid-Luttinger liquid resistance $R_C = \frac{h}{e^2}\frac{K - 1}{2K}$ for AT fixed point (other fixed points are studied analogously). In red are the voltages for the non-trivial case of grounding the superconducting lead and one of the normal leads.
}\label{fig:network}
\end{figure}

Let us now transform the conductance matrices above into the resistance networks. Under the assumption that there is no backscattering at the Fermi liquid-Luttinger liquid interface, we can treat it as a resistance $R_C = \frac{h}{e^2}\frac{K - 1}{2K}$ added into the network.\cite{Cha03} This assumption is valid for adiabatic contacts.\cite{Sed12, Sed14}

For the normal reflection case, obviously, resistances between all of the contacts are infinite and adding the contact resistance does not change the picture.

For the AR case there is resistance $R_{AR} = \frac{h}{e^2} \frac{1}{2K}$ between each of the normal contacts and the superconductor. In series with the $R_C$ the total resistance is $\frac{h}{e^2}$, i.e. the non-interacting answer.

For the AT case the resistance between each of the normal leads and the superconductor is $\frac{h}{2e^2}\frac{1}{K}$, and between the normal leads it is $-\frac{h}{e^2}\frac{1}{K}$. Let us now consider a few voltage setups to check that for them the conductance is the same as in the non-interacting cases. Firstly, we assume both normal leads are at the same voltage, superconductor at a different one. Then the total resistance is $\frac{h}{e^2}$, like in the non-interacting case. Secondly, we assume one of the normal leads and the superconductor are at the same voltage and the other normal lead is at different one. This situation is shown in Fig. \ref{fig:network}. Let us make an explicit calculation for that case. The current conservation conditions read:
\begin{align}
\frac{2 K}{K-1} (V - V_1) = 2 K V_1 + K (V_2 - V_1), \\
K (V_2 - V_1) = 2 K V_2 + \frac{2 K}{K - 1} V_2.
\end{align}
Here we dropped the trivial common factor $e^2/h$. These equations have the solution:
\begin{align}
V_2 = - \frac{K - 1}{K + 1} V_1, \\
V_1 = \frac{K + 1}{2K}V.
\end{align}
Then the total current from the biased lead into the system is:
\begin{align}
I = (V - V_1) \frac{e^2}{h} \frac{2K}{K - 1} = \frac{e^2}{h}V.
\end{align}
Again, as expected, the $K$ factor is dropped from the expression. These results imply the familiar expressions for the conductances:
\begin{align}
G_{11} = G_{12} = G_{21} = G_{22} = \frac{e^2}{h},
\end{align}
same ones as obtained for the non-interacting leads.

Note that all the above results are only valid when there is substantial decoherence inside the Luttinger liquid, i.e. when the length of the LL part is larger than $v_F \tau_d$, where $\tau_d$ is the decoherence time. The Luttinger liquid part should  also be long  compared to the Majorana screening cloud (similar to the Kondo cloud), the crossover length between the high-energy fixed point and the low-energy one.\cite{Aff14}

\section{Conclusion and discussion\label{sec:conclusion}}

We have discussed a Kramers pair of Majoranas in contact with a spinless Luttinger liquid. The most promising setups to realize the system include a Josephson junction with phase difference $\pi$ between the superconductors on top of the QSH system. Examples are shown in Fig. \ref{fig:setup}. 

We find that in the non-interacting case at zero energy the system realizes  Andreev reflection, Andreev transmission, or a combination of the two. These are two types of Andreev processes. In the first the reflected hole goes into the same lead the electron arrived from, while in the second the hole goes into a different lead. Which combination is realized depends on microscopic details. In the special $U(1)$-symmetric case, only Andreev reflection can occur. 

When we take interactions into consideration, we find that, for repulsive interactions of moderate strength, only Andreev transmission is stable. 
For strong repulsive interactions or strong local impurity-quantum wire interactions with $U(1)$ symmetry, normal transmission, as well as Andreev reflection, becomes stable. Depending on the bulk interactions in the Luttinger liquid, either only trivial critical points occur for $K>1/4$, or a non-trivial critical point emerges for $K<1/4$. 

When the $U(1)$ symmetry is broken, for strong interactions $K<1/2$ normal reflection (with spontaneously breaking time-reversal symmetry) becomes stable. This leads to the surface separating flows to the normal reflection and Andreev transmission fixed points. We know that at $K<1/4$ the surface includes non-trivial fixed points in $U(1)$ symmetric cases and the flow along the surface is towards them. The surface mostly lies in strong-coupling regime and therefore the full analysis and proof of absence of additional structure on it is beyond the scope of the current manuscript.

Finally, we discussed the transport signatures, in particular conductance at all the trivial fixed points. Computing conductance at non-trivial ones requires further study. 

This brings us to the outlook for future study. Further studies of the $U(1)$ breaking case are necessary to substantiate the picture of the present manuscript. Numerical study of the model would be very useful, checking our predicted phase diagram and in particular the non-trivial critical point. 

The questions we addressed in the present paper are only the first ones in a series of questions on how  interactions in the leads influence the observation of  non-trivial topology in the new symmetry classes. So far only class D is discussed in enough details;\cite{Fid12} we expect much interesting physics to occur in BDI, CII symmetry classes, where more than 1 Majorana per edge of the topological superconductor is possible.

\textit{Note added.} When this manuscript was almost finished,  a preprint appeared on the same system, albeit without taking into account  interactions. \cite{JLi15}

\acknowledgements
We thank J. Folk, R. Lutchyn and D. Liu for stimulating discussions. The research of DIP and YK was supported by the UBC-Max Planck Centre and NSERC.  The research of IA was supported by NSERC Discovery Grant 36318-2009 and CIFAR.

\appendix
\renewcommand\theequation{\Alph{section}.\arabic{equation}}

\section{Bosonization }

In this section we give a pedagogical introduction into the bosonisation conventions we use. We confirm the formulas by comparing of the finite size spectrum of antiperiodic Luttinger liquid using the fermionic and bosonic representations. We start by introducing left- and right-moving bosons $\phi_{R, L}$ according to: 
\be \psi_{R,L}=\Gamma e^{i\sqrt{4\pi}\phi_{R/L}}.\ee
We require
\be \{\psi_R(x),\psi_L(y)\}=0.\ee
Thus
\be e^{i\sqrt{4\pi}\phi_R(x)}e^{i\sqrt{4\pi}\phi_L(y)}=-e^{i\sqrt{4\pi}\phi_L(y)}e^{i\sqrt{4\pi}\phi_R(x)}.
\ee
Using
\be e^Ae^B=e^{A+B+(1/2)[A,B]}\ee
\be e^{-2\pi [\phi_R(x),\phi_L(y)]}=-e^{2\pi [\phi_R(x),\phi_L(y)]}
\ee
or
\be e^{4\pi [\phi_R(x),\phi_L(y)]}=-1.\ee
Thus, we see that $\phi_R$ and $\phi_L$ don't commute with each other but their commutator is a constant. We can choose it to be:
\be  [\phi_R(x),\phi_L(y)]=i/4.\ee
($-i/4$ would work equally well.) We also require
\be \{\psi_{R/L}(x),\psi_{R/L}(y)\}=0\ee
implying
\be e^{4\pi [\phi_{R/L}(x),\phi_{R/L}(y)]}=-1.\ee
We can satisfy these conditions by
\be [\phi_{R/L}(x),\phi_{R/L}(y)]=\pm {i\over 4}\epsilon (x-y)
\ee
Using
\bea \phi &\equiv& \phi_R+\phi_L\nonumber \\
\theta &\equiv &\phi_L-\phi_R,\eea
we see that:
\be [\phi (x),\phi (y)]={i\over 4}[\epsilon (x-y)-\epsilon (y-x)+1-1]=0,\ee
Similarly, $[\theta (x),\theta (y)]=0.$ 
\be [\phi (x),\theta (y)]={i\over 4}[\epsilon (x-y)+\epsilon (x-y)-1-1]=-i\Theta(y-x).
\ee
where $\Theta(x)$ is the Heaviside theta-function (=1 for $x>0$ and =0 for $x<0$). We now write
\be \psi_{L/R}=\Gamma e^{i\sqrt{\pi}[\phi \pm \theta ]}.\ee

Now consider a fermonic chain with open boundary conditions defined on sites $j=1, 2, \ldots \ell -1$ with $\ell$ odd so that the number of sites is even. 
Assume we are at half-filling and use
\be c_j\approx i^j\psi_R(j)+(-i)^j\psi_L(j).\ee
Then the phantom site boundary conditions, $c_0=c_{\ell}=0$ imply
\be \psi_R(0)=-\psi_L(0),\ee
\be \psi_R(\ell )=\psi_L(\ell ).\ee
Let's see what boundary conditions these imply on the boson fields.  In terms of right and left movers, these are
\be e^{i\sqrt{4\pi}\phi_R(0)}=-e^{i\sqrt{4\pi}\phi_L(0)}\label{0}\ee
\be e^{i\sqrt{4\pi}\phi_R(\ell )}=e^{i\sqrt{4\pi}\phi_L(\ell )}.\label{ell}\ee
Choosing $\delta$ to be an infinitesimal positive number, the first equation implies:
\be e^{i\sqrt{4\pi}\phi_R(0)}e^{-i\sqrt{4\pi}\phi_R(\delta )}=-e^{i\sqrt{4\pi}\phi_L(0)}e^{-i\sqrt{4\pi}\phi_R(\delta )}
\ee
\begin{align} e^{i\sqrt{4\pi}[\phi_R(0)-\phi_R(\delta )]}e^{2\pi [\phi_R(0),\phi_R(\delta )]}\nonumber \\=-e^{i\sqrt{4\pi}[\phi_L(0)-\phi_R(\delta )]}e^{2\pi [\phi_L(0),\phi_R(\delta )]}
\end{align}
Assuming $\phi_R(x)$ to be continuous, and taking $\delta \to 0^+$, we obtain:
\be e^{-i\pi /2}=-e^{i\sqrt{4\pi}[\phi_L(0)-\phi_R(0)]}e^{-i\pi /2}.
\ee
Thus
\be e^{i\sqrt{4\pi}[\phi_L(0)-\phi_R(0)]}=-1\ee
implying
\be \phi_L(0)-\phi_R(0)=\sqrt{\pi }/2,\ \  (\hbox{mod}\ \sqrt{\pi}).
\ee
On the other hand, Eq. (\ref{ell}) implies
\be e^{i\sqrt{4\pi}\phi_R(\ell )}e^{-i\sqrt{4\pi}\phi_R(\ell -\delta )}=e^{i\sqrt{4\pi}\phi_L(\ell )}e^{-i\sqrt{4\pi}\phi_R(\ell -\delta )}
\ee
where again $\delta$ is a positive infinitesimal. Thus
\begin{align} e^{i\sqrt{4\pi}[\phi_R(\ell )-\phi_R(\ell -\delta )]}e^{2\pi [\phi_R(\ell ),\phi_R(\ell -\delta )]}\nonumber \\ =e^{i\sqrt{4\pi}[\phi_L(\ell )-\phi_R(\ell -\delta )]}
e^{2\pi [\phi_L(\ell ),\phi_R(\ell -\delta )]}.
\end{align}
Now taking $\delta \to 0^+$, 
\be e^{i\pi /2}=e^{i\sqrt{4\pi}[\phi_L(\ell )-\phi_R(\ell  )]}e^{-i\pi /2}.
\ee
Again we find
\be e^{i\sqrt{4\pi}[\phi_L(\ell )-\phi_R(\ell )]}=-1\ee
implying 
\be \phi_L(\ell )-\phi_R(\ell )=\sqrt{\pi }/2,\ \  (\hbox{mod}\ \sqrt{\pi}).
\ee
Remarkably, while the boundary conditions on the $\psi_{R/L}$ have opposite sign at $x=0$ and $x=\ell$, the boundary conditions on the 
bosons are the same at $x=0$ and $x=\ell$. 

These boundary conditions on the bosons can be seen to correctly reproduce the finite size spectrum implied by the boundary conditions 
on the fermions.  
This follows since the mode expansion is
\be \phi (x)=\sqrt{\pi}\left({1\over 2}+{xn\over \ell}\right) +\ldots \ee
yielding
\be E={\pi n^2\over 2\ell}+\ldots\label{Eb}\ee
for non-negative integer $n$.  The boundary conditions on the fermions are equivalent to right-movers only 
with anti-periodic boundary conditions on an interval of length $2\ell$ and hence wave-vectors
\be k_m={\pi (2m+1)\over 2\ell}.\ee
The energy of the lowest energy state with $n$ fermions added to the ground state is hence
\be E_n={\pi \over 2\ell}\sum_{m=0}^{n-1}(2m+1)={\pi n^2\over 2\ell}\ee
in agreement with Eq. (\ref{Eb}). We have thus obtained the bosonisation formulas used in the main text as well as confirmed them by explicit comparison of the finite size spectrum of the fermionic and bosonic formulations of the problem.

Noting that these boundary conditions are equivalent  to 
\be \theta (0)=\theta (\ell )=\sqrt{\pi }/2,\ \  (\hbox{mod}\ \sqrt{\pi}), \ee
we see that
\be \psi_{R/L}(0)=\pm i\Gamma e^{i\sqrt{\pi}\phi (0)},\ \  \psi_{R/L}(\ell )=\pm i\Gamma e^{i\sqrt{\pi}\phi (\ell )}.
\ee
The sign should be opposite for left and right movers but the overall sign, and whether it is the same at $x=0$ and $x=\ell$, seems ambiguous. 

Now consider the Majorana couplings:
\be H_b=-it_L\gamma_L(c_1^\dagger +c_1)/2-t_R\gamma_R(c_{\ell -1}^\dagger -c_{\ell -1})/2.\ee
We may use
\be c_1=i(\psi_R-\psi_L)=2i\psi_R(0),\ \  c_{\ell -1}=\psi_R+\psi_L=2\psi_R(\ell ),\ee
[assuming ($\ell -1)/2$ is even] yielding
\begin{align} H_b=t_L\gamma_L[\psi_R(0)-\psi_R^\dagger (0)]-t_R\gamma_RL[\psi_R(\ell )-\psi_R^\dagger (\ell )]\nonumber \\ =\pm it_L\gamma_L\Gamma \cos {\sqrt{\pi}\phi (0)}
\pm it_R\gamma_R\Gamma \cos {\sqrt{\pi}\phi (\ell )}.
\end{align}
Note that both boundary terms involve the cosine, not one cosine and one sine. The sign of each boundary term can be 
changed by changing the signs of $\gamma_L$ and $\gamma_R$. So these signs are not important and can  both  be  chosen to be positive.

\section{Formation of Majorana bound states}
\label{App:MBS}
Imagine in either of the setups shown in Fig. \ref{fig:setup}a,b that the tunneling between the Josephson junction and the outside leads is set to zero. Let us find the bound states in this case. This is the well-known problem of a Josephson junction on top of a QSH edge.\cite{LFu09, Bee12} The equation for Andreev bound states then reads:
\begin{align}
{\rm Det}\left[1 - \alpha^2(E) \Lambda s(E) \Lambda^* s^*(-E)\right] = 0.
\end{align}
Here $\alpha$ is the energy-dependent amplitude of Andreev reflection, $\alpha(E\ll \Delta) \approx 1$, $s(E)$ is the normal region scattering matrix written in the basis of the electron states near the left superconductor and the right one ($s^*(E)$ is correspondingly the scattering matrix in the basis of the hole states near the left superconductor and hole states near the right one), and $\Lambda$ is the matrix of the reflection phases between electrons and holes near the left and right superconductors. For the superconducting phases $\phi_1$ and $\phi_2$ mode-matching and locality of Andreev reflection gives $\Lambda = {\rm diag}\{e^{i \phi_1}, - e^{i\phi_2}\}$. Notice the minus sign due to opposite spins of the helical electrons. As the middle part of the junction preserves time-reversal symmetry, the normal scattering matrix has only transmission:
\begin{align}
s(0) = \begin{pmatrix}
0 & e^{i\chi} \\ e^{i \chi} & 0
\end{pmatrix}.
\end{align}
The condition for the bound states to be at zero then is equivalent to requiring eigenvalues of the matrix
\begin{align}
\Lambda s(0) \Lambda^* s^*(0) = \begin{pmatrix}
- e^{i (\phi_1 - \phi_2)} & 0 \\
0 & - e^{- i (\phi_1 - \phi_2)}
\end{pmatrix}
\end{align}
to be $1$. Thus:
\begin{align}
\phi_1 - \phi_2 = \pi.
\end{align}

\section{Single-lead Model}
As a side problem we study the single-lead case. In this case we have 1 normal wire on the $x>0$ axis interacting with two Majorana modes at the origin. 
Before coupling to the Majorana mode's we impose a normal scattering boundary condition at the end of the wire:
\be \psi_L(0)=\psi_R(0).\label{NRbcSL}\ee
Thus there is only one independent field at $x=0$, which we simply denote by $\psi (0)$,  and the tunnelling term simplifies to:
\be H_T=d^\dagger [t_1\psi (0)+t_2\psi^\dagger (0)]+h.c.\ee
We can again choose  $t_1$ and $t_2$ real by redefining the phases of $\psi$ and $d$. This model  has a particle-hole symmetry forbidding a $d^\dagger d$ term:
\bea \psi_{R/L}&\to& \psi_{R/L}^\dagger \nonumber \\
d&\to& -d^\dagger .\eea
Note that this symmetry also forbids a $\psi^\dagger (0)\psi (0)$ term but allows the term
\be H_U\equiv U(d^\dagger d-1/2):\psi^\dagger (0)\psi (0):.\ee
which we will consider. 

The solution for the non-interacting system is trivial. 
	For one-channel scattering matrix only two possibilities at zero energy are realized: perfect normal and perfect Andreev reflection with possible phase.\cite{Ber09} This is a consequence of  S-matrix unitarity and  particle-hole symmetry. Moreover, we notice that the total phase of the S-matrix at zero energy is fixed due to the number of (quasi-)bound states at the junction. Here we have two Majorana bound states, and when both of them are coupled to the lead the scattering matrix is fixed to be completely normal-reflecting with phase shift $\pi$. The constraint on the single-channel S-matrix overcomes the natural tendency towards Andreev reflection in the NS junctions, which is present in the 2-channel case above. 
	
\begin{figure}[t]
\centering
\includegraphics[width = \linewidth]{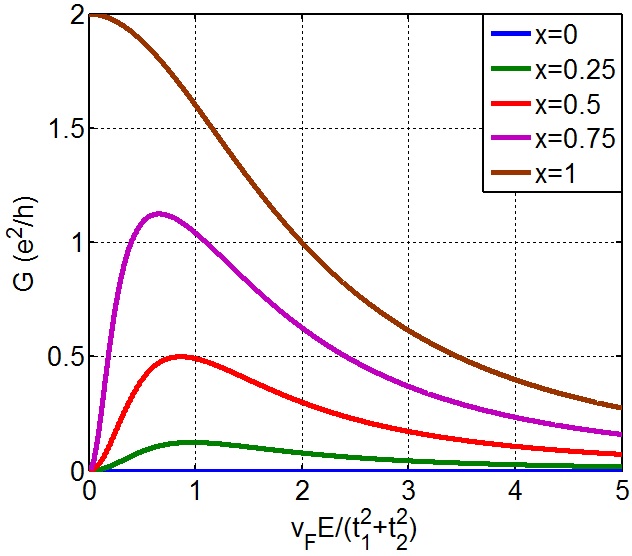}
\caption{Conductance of a single non-interacting lead in contact with two Majorana fermions. The coupling asymmetry parameter $x=2t_1t_2/(t_1^2 + t_2^2)$ takes values $0..1$ for different curves.}\label{fig:Fig8}
\end{figure}

Explicitly solving the BdG equations, we get to
\begin{align}
iv_FE(r_N+r_A-1)&=- (t_1 - t_2)^2[(1+r_N)+r_A],\\
iv_FE(r_N-r_A-1)&=- (t_1 + t_2)^2[(1+r_N)-r_A];
\end{align}
where $r_A$ and $r_N$ are the amplitudes of Andreev and normal reflection correspondingly. We can use these results together with known BTK formula for the single channel -- $G(E)=2\abs{r_A}^2$ -- to write
\be
G(\tilde E)={2}{x}\Big[\frac{(1+x)^2}{\tilde E^2+(1+x)^2}-\frac{(1-x)^2}{\tilde E^2+(1-x)^2}\Big]
\ee
for the zero-temperature differential conductance where
\be
x=\frac{2 t_1 t_2}{t_1^2+t_2^2}, \qquad \tilde E=\frac{v_F E}{t_1^2+t_2^2}.
\ee
The Andreev (differential) conductance as a function of bias voltage is plotted in Fig.\,\pref{fig:Fig8}.
The conductance maximum happens at $\tilde E=\sqrt{1-x^2}$ and it is $G=2x^2\times e^2/h$.
At zero energy, when $t_1 \neq t_2$, and both the Majoranas are coupled to the lead, the system exhibit full normal reflection with a phase shift $\pi$. 

To treat the interacting case it is again convenient to bosonize.  The normal reflection boundary condition of Eq. (\ref{NRbcSL}) implies that $\theta (0)=0$ so 
\be H_T \to d^\dagger \Gamma \left[ \tilde t_1e^{i\sqrt{\pi}\phi (0)}+\tilde t_2e^{-i\sqrt{\pi}\phi (0)}\right] +h.c. \ee
where $\tilde t_1$ and $\tilde t_2$ are rescaled $t_1$ and $t_2$ parameters. 
In terms of the  two Majorana fields, defined by $d=(\gamma_1+i\gamma_2)/2$, this becomes:
\be H_T=(\tilde t_1-\tilde t_2)i\gamma_1\Gamma \sin \sqrt{\pi}\phi (0)-(\tilde t_1+\tilde t_2)i\gamma_2\Gamma \cos \sqrt{\pi}\phi (0)\ee
The bosonized form of $H_U$ is
\be H_U=\tilde Ui\gamma_1\gamma_2\partial_x \theta (0).\ee
This is actually equivalent to (the spin sector of) a Kondo model with an interacting lead as can be seen by the exact mapping
to spin operators:
\begin{align}
S^y = i \gamma_1 \Gamma, \; S^x =- i \Gamma \gamma_2,\; S^z=i \gamma_2 \gamma_1,
\end{align}
The spin-up/down states are the filled/empty charge states of the $d$-level. To make the mapping to the Kondo model more transparent it is to convenient to define  an effective spin boson by the canonical transformation:
\bea \phi_s&=&\phi /\sqrt{2}\nonumber \\
\theta_s&\equiv& \sqrt{2}\theta .\eea
Then
\bea &&H_T+H_u =(\tilde t_1-\tilde t_2)S^y\sin \sqrt{2\pi}\phi_s(0)\nonumber \\
&&+ (\tilde t_1+\tilde t_2)S^x\cos \sqrt{2\pi}\phi_s(0) +(\tilde U/\sqrt{2})S^z\partial_x\theta_s(0).\nonumber \\
&& \eea
For free spinful fermions at a boundary with $\psi_{L\alpha}(0)=\psi_{R\alpha}(0)$, abelian bosonization in terms of spin and charge bosons gives:
\bea \psi^\dagger \sigma^x\psi (0)&\propto& \cos \sqrt{2\pi}\phi_s(0)\nonumber \\
\psi^\dagger \sigma^y\psi (0)&\propto& \sin \sqrt{2\pi}\phi_s(0)\nonumber \\
\psi^\dagger  \sigma^z\psi (0)&\propto& \partial_x\theta_s(0)\eea
with each operator having dimension 1.
Thus
\be H_T+H_U=\sum_{i}J_iS^is^i(0)\ee
where $\vec s(x)$ is the conduction electron spin density. The Luttinger parameter in the spin boson formulation is $K_s=2K$ 
corresponding to strong attractive interactions which lower the scaling dimension of the boundary spin density operators $s^x$ and $s^y$ 
from 1 to 1/2 in the case $K=1$.  The Kondo model with an interacting lead is well understood. The weak coupling renormalisation group equations, 
describing the flow of the effective coupling constants with energy, are:
\begin{align}
\frac{d J_z}{d\ell} &= \nu J_xJ_y \\
\frac{d J_x}{d\ell} &=[1-1/(2K)]J_x+\nu J_y J_z\\
\frac{d J_y}{d\ell} &=[1-1/(2K)]J_y+\nu J_x J_z
\end{align}
where $\nu=1/(2\pi v_F)$ is the density of states at the Fermi energy and $\ell \equiv -\log D$ where $D$ is a UV cut-off energy scale. 

The solution of these RG equations is well-known. All 3 couplings grow to large values, resulting in the Kondo infrared fixed point which is characterized 
by the screening of the impurity spin and a $\pi /2$ phase shift for the conduction electrons. 
In the original fermionic language this fixed point corresponds to a fermion from the lead coupling strongly with the two Majoranas, sharing an electron 
in an entangled state while the low energy electrons undergo purely normal reflection with a $\pi /2$ phase shift. 

Notice that the special case of the non-interacting model when $t_1=t_2$ and the system exhibit Andreev reflection at zero bias corresponds to the special set of initial conditions for the RG equations: $J_y^0\neq 0$, while $J_x^0=J_z^0=0$. For these conditions no $J_x$ or $J_z$ term gets generated in RG. The results in this case coincide with the single-Majorana case,\cite{Fid12} i.e. for $K>1/2$ $J_y\to\infty$, corresponding to the Andreev reflection fixed point, and for $K<1/2$ $J_y\to 0$, corresponding to normal reflection without phase shift. Either of the fixed points is unstable since for non-zero $J_x^0$ or $J_z^0$ both $J_x$ and $J_z$ get generated and flow to large couplings simultaneously, bringing the system to the stable Kondo fixed point with normal reflection and $\pi/2$ phase shift.

These results coincide with the non-interacting results from above. 

Note that the time reversal symmetry of the 2-channel model, Eq. (2.3),
takes $\gamma_1\to - \gamma_2$, $\gamma_2\to \gamma_1$ and thus forbids a
decoupling of one of the Majoranas. On the other hand, the particle-hole
symmetry of the single channel model, Eq. (B.3), takes $\gamma_1\to
-\gamma_1$, $\gamma_2\to \gamma_2$, and thus {\it is} consistent with a
decoupling of one of the Majoranas, leading to the critical point
discussed earlier in this Appendix which is not possible in the time
reversal symmetric 2 channel case.

\section{Tight-binding model}

It is possible to obtain the continuum model studied above as the low energy limit of a tight-binding model, which is convenient for some purposes 
including numerical simulations.  We may obtain the $\tilde \psi_{R/L}$ operators, occurring after the $C_L$ transformation,  as the continuum limit of tight-binding operators at half-filling:
\be c_n\approx i^n\tilde \psi_R(n)+(-i)^n\tilde \psi_L(n)\ee
where $\tilde \psi_{R/L}(x)$ vary slowly on the lattice scale. The CT transformation of the $C_L$ transformed continuum model, of Eq. (\ref{CT}), corresponds to
\bea c_n&\to& i(-1)^nc_n^\dagger \nonumber \\
d&\to& -id^\dagger \nonumber \\
i&\to& -i.\label{CTL}
\eea
The bulk terms in the Hamiltonian are those of the standard spinless interacting tight-binding model:
\begin{align} 
H_0+H_{\hbox{int}}=\sum_n&\left[-t(c_n^\dagger c_{n+1}+h.c.)\right.\nonumber \\ &\left.+(V/4)(c^\dagger_nc_n-1/2)(c^\dagger_{n+1}c_{n+1}-1/2)\right]
\end{align}
The various boundary terms are obtained as follows:
\begin{align} t_1d^\dagger c_0 \to& t_1d^\dagger [\tilde \psi_R(0)+\tilde \psi_L(0)]\nonumber \\
-it_2d^\dagger (c_1^\dagger -c_{-1}^\dagger )/2 \to& t_2d^\dagger [\tilde \psi_L^\dagger (0)-\tilde \psi_R ^\dagger (0)]\nonumber \\
U_1i[c_1^\dagger-c_{-1}^\dagger )c_0/2-h.c.]\to&\nonumber \\ \to U_1[\tilde \psi_R^\dagger (0)\tilde \psi_R(0)&-\tilde \psi_L^\dagger (0)\tilde \psi_L(0)]\nonumber \\
U_2[c_0^\dagger (c_1-c_{-1})/2+h.c.]\to&\nonumber \\ \to iU_2[\tilde \psi_R^\dagger (0)\tilde \psi_L(0)&-\tilde \psi_L^\dagger (0)\tilde \psi_R(0)],
\end{align}
all of which can be seen to be symmetric under the CT symmetry of Eq. (\ref{CTL}). The $U(1)$ symmetry is also evident if either $t_1$ or $t_2=0$. 

\section{Partition function and impurity entropy \label{app:Z}}
The partition function of our system is expected to have the form:
\be Z=ge^{\pi LT/(6u)}\ee
in the limit
\be {u\over L}\ll T\ll T_K,D\ee
where $T_K$ is the crossover scale and $D$ is the bandwidth. $g$ is a universal number whose logarithm gives the impurity entropy.  The $g$-theorem implies that the parameter $g$ decreases under RG flow or should have a constant value on a line of fixed points connected by a marginal operator.\cite{Aff91} We calculate it for 
anti-periodic boundary conditions on the fermions on a chain of length $L$, 
\be \psi_{R/L}(L)=-\psi_{R/L}(0)\ee
at the ($t_i=0$) normal Transmission, Andreev Reflection and Andreev Transmission fixed points.

\subsection{Normal Transmission fixed point}
In this case we  have an anti-periodic chain with a decoupled $d$-level so $g$ simply equals $2$ from the 2 states of the $d$-level, for any Luttinger parameter $K$.  For an explicit 
proof that there is no contribution to $g$ from the anti-periodic chain, see Appendix A and [\onlinecite{Eggert92}].

\subsection{Andreev Reflection Fixed Point}
Recalling that Andreev Reflection corresponds to normal reflection after the $C_L$-transformion, we now have the boundary conditions:
\bea \tilde \psi_R(0)&=&\tilde \psi_L(0)\nonumber \\
\tilde \psi_R(L)&=&\tilde \psi_L(L).\label{eq:ARBC}\eea 
For the non-interacting case, we may make an ``unfolding'' transformation, defining the system with right-movers only on an interval of length $2L$ with 
periodic boundary conditions by
\be \tilde \psi_R(-x)=\tilde \psi_L(x),\ \  0\leq x\leq L.\label{NRbc}\ee
The right-movers have allowed wave-vectors
\be k={\pi n\over L}\ee
leading to the partition function
\be Z=2\left\{\prod_{n=1}^\infty \left[1+e^{-\pi v_F n/(LT)}\right]\right\}^2.\label{ZARFF}\ee
(The square arises from equal contributions from particles and holes, and 2  from the zero mode for $n=0$.) For large $LT/v_F$. Using the Euler-Maclaurin expansion:
\be \sum_{n=1}^\infty f(n)\approx \int_0^\infty dn f(n)-(1/2)f(0)+O(f'(0))\label{EM1}\ee
we see that
\be \ln Z\approx \ln 2 +2{LT\over \pi v_F}\int_0^\infty dx \ln \left( 1+e^{-x}\right)-\ln 2={\pi LT\over 6v_F}
\ee
giving 
\be g=1.\ee
For the interacting case we bosonize.  The boundary conditions of Eq. (\ref{NRbc}) imply
 \bea \theta (0)&=&0\nonumber \\
 \theta (L)&=&\sqrt{\pi}/2 \ \  (\hbox{mod}\ \sqrt{\pi}).\eea
This leads to the mode expansion:
\bea \theta (x)&=&{\sqrt{\pi}(Q+1/2)x\over L}+i\sum_{n=1}^\infty\sqrt{\tilde K\over \pi n} \sin (\pi nx/L)(a_n-a_n^\dagger ) \nonumber \\
\phi (x)&=&\sum_{n=1}^\infty {1\over \sqrt{\pi \tilde Kn}}\cos  (\pi nx/L)(a_n+a_n^\dagger ).
\eea
for integer $Q$ and harmonic oscillator lowering operators $a_n$. The corresponding finite size spectrum is:
\be E+{\pi u\over L}\left[{(Q+1/2)^2\over 2\tilde K}+\sum_{n=1}^\infty m_nn\right]\ee
for non-negative harmonic oscillator quantum numbers, $m_n$. 
The corresponding partition function:
 \begin{align} 
 Z=\sum_{Q=-\infty}^\infty e^{-\pi u(Q+1/2)^2/(2LT\tilde K)}\prod_{n=1}^\infty \left[1-e^{-\pi un/(LT)}\right]^{-1}.
 \end{align}
 For $\tilde K=1$ this can be shown to be the same as Eq. (\ref{ZARFF}) using the Jacoby triple product identity.  For large $LT/u$, 
\bea &&\sum_{Q=-\infty}^\infty e^{-\pi u(Q+1/2)^2/(2LT\tilde K)}\nonumber \\
&&\approx \int_{-\infty}^\infty dQ e^{-\pi uQ^2/(2LT\tilde K)}=\sqrt{2LT\tilde K\over u}
\eea
and, using a Dedekind eta-function identity:
\be \prod_{n=1}^\infty \left[1-e^{-\pi un/(LT)}\right]^{-1}\approx \sqrt{u\over 2LT}e^{\pi LT/(6u)}.
\ee
 Thus we see that 
 \be g=\sqrt{\tilde K}=1/\sqrt{K}.\ee

 \subsection{Andeev Transmission Fixed Point}
Now the boundary conditions become:
\bea \psi_R(0)&=&-\psi_R^\dagger (L)\nonumber \\
\psi_L(0)&=&\psi_L^\dagger (L),\label{bcAT}\eea
which are consistent with time-reversal symmetry. 
For the non-interacting case it is convenient to decompose $\psi_{R/L}$ into Hermitean and anti-Hermitean parts:
\be \psi_{R/L}=(\chi_{R/L}+i\chi_{R/L}')/2.\ee
We then see that $\chi_{R}'$, $\chi_L$ obey periodic boundary conditions while $\chi_R$, $\chi_L'$ obey anti-periodic boundary conditions on the interval of length $L$. 
The corresponding energies are thus
\bea E^{R'}_n&=&{\pi v_F\over L}2n \ \  (n=1,2, \ldots )\nonumber \\
E^{R}_n&=&{\pi v_F\over L}(2n+1),\ \  (n=0,1,2, \ldots )\label{fssAT}
\eea
with an identical spectrum for $\chi_L$,  $\chi_L'$. 
In addition there  are two zero energy Majorana modes, corresponding to $\chi_R'(x)$, $\chi_L(x)$ being constant.  These can be combined 
to make one zero energy Dirac fermion   operator.  Thus the partition function is:
\be Z=2\left\{ \prod_{n=1}^\infty \left[1+e^{-\pi v_Fn/(LT)}\right]\right\}^2\label{ZAT}\ee
with the factor of $2$ arising due to the zero energy mode. This is the same as in \eqref{ZARFF}.

To treat the interacting case, we again bosonize. Now the boundary conditions of Eq. (\ref{bcAT}) imply
\bea \phi (L)+\phi (0)&=&-\sqrt{\pi}  \ \  (\hbox{mod} \ 2\sqrt{\pi})\nonumber \\
\theta (L)+\theta (0)&=&\sqrt{\pi}  \ (\hbox{mod}\  2\sqrt{\pi})\label{bcATb}\eea
The corresponding mode expansion is:
\begin{align} 
&\phi (x)=\phi_0+\sum_{n=0}^\infty {1\over \sqrt{2K\pi (2n+1)}}
\nonumber \\ &\times \left[ e^{i\pi (2n+1)x/L}a_{2n+1,R}+e^{-i\pi (2n+1)x/L}a_{2n+1,L}+h.c. \right] \nonumber \\
 &\theta (x)=\theta_0+\sum_{n=0}^\infty \sqrt{K\over 2\pi (2n+1)}\nonumber \\ &\times \left[ e^{i\pi (2n+1)x/L}a_{2n+1,R}-e^{-i\pi (2n+1)x/L}a_{2n+1,L}+h.c. \right]
\label{modeAT}
\end{align}
where there are two possible inequivalent choices for the constant terms: $(\phi_0,\theta_0)=\sqrt{\pi}(1,-1)/2$ or $\sqrt{\pi}(-1,1)/2$.  $a_{nR}$ and $a_{nL}$ are 
independent harmonic oscillator annihilation operators for right and left movers. 
This yields the partition function
\be Z=2\prod_{n=0}^\infty \left[1-e^{-\pi (2n+1)u\beta /L} \right]^{-2}\label{ZATb}
\ee
where the factor of $2$ arises from the two  choices of $(\phi_0,\theta_0)$. Using:
\begin{align} \prod_{n=0}^\infty &(1-q^{2n+1})^{-1}={\prod_{n=1}^\infty (1-q^{2n})\over \prod_{n=1}^\infty (1-q^{n})}\nonumber \\ &=\prod_{n=1}^\infty { (1-q^n)(1+q^n)\over 1-q^n}
=\prod_{n=1}^\infty (1+q^n)
\end{align}
we see that the two partition functions Eq. (\ref{ZAT}) and (\ref{ZATb}) are identical. The mode expansion of Eq. (\ref{modeAT}) is independent of 
interactions, parameterised by $K$, since the boundary conditions of Eq. (\ref{bcATb}) do not permit any winding modes. 
Thus we conclude that $g=1$ at the AT fixed point, for all $K$. 

These results are all consistent with the RG flows discussed in Sec. III. For $1/4<K<1$ an RG flow from the NT to AR fixed point is allowed by the $g$-theorem. 
For all $K$ an RG flow from the NT to AT fixed point is allowed. For all $K<1$, an RG flow from the AR to AT fixed point is allowed. 

\subsection{Partial Andreev Reflection and Partial Andreev Transmission: Non-Interacting Case}

 We will now prove that $g=1$ along the entire line of fixed points in the 
non-interacting case characterised by partial Andreev reflection and partial Andreev transmission. 
For this we use the scattering matrix obtained in Eq. (\ref{eq:S}):
\begin{align}
S_{eh} = \frac{1}{t_1^2 + t_2^2}\begin{pmatrix}
-2 t_1 t_2 &  t_1^2 - t_2^2\\
t_1^2 - t_2^2 & 2 t_1 t_2
\end{pmatrix} =  S_{he}.
\end{align}
This connects:
\begin{align}
\begin{pmatrix}
\psi_R^\dag(0)\\
\psi_L^\dag(L)
\end{pmatrix} = S_{eh} \begin{pmatrix}
\psi_R(L)\\
\psi_L(0)
\end{pmatrix},\\
\begin{pmatrix}
\psi_R(0)\\
\psi_L(L)
\end{pmatrix} = S_{he} \begin{pmatrix}
\psi_R(L)^\dag\\
\psi_L(0)^\dag
\end{pmatrix}.
\end{align}
On top of this scattering, the evolution of the wavefunction along the closed ring is described by  rotation by an angle $k L$ due to the evolution above/below the Fermi energy. Notice that the angle is the same for electrons and holes, as when we go from one to another the momentum changes sign, the exponent must be complex conjugated as well. Besides, the angle is the same for left- and right-moving electrons, since when we apply the spatial parity transformation, again $k\to -k$ and $x\to -x$. Combining the steps of the evolution: Andreev reflection, evolution along the ring, Andreev reflection, evolution along the ring, we obtain the following condition on the eigenstates and consequently eigenmomenta:
\begin{align}
\begin{pmatrix}
\psi_R(2L)\\
\psi_L(2L)
\end{pmatrix} =S_{eh}^2\begin{pmatrix}
\psi_R(0)\\
\psi_L(0)
\end{pmatrix},\\
\end{align}
Since $S_{eh}^2={\bf I}$, we obtain the allowed wave-vectors $\pi n/L$ and the same partition function as in Eq. (\ref{ZARFF}), for all $t_1/t_2$.  
This constant value of $g$ is consistent with a line of fixed points between AR and AT for the non-interacting cases, as is implied by the S-matrix.

\subsection{Normal Reflection Fixed Point}

Finally, the calculation of the section on Andreev Reflection critical point can be directly applied to the Normal Reflection one. In the AR section we computed the impurity entropy of the NR critical point in the $C_L$-transformed model, see eqs. \eqref{eq:ARBC}. In the original the NR boundary conditions can have an additional phase of the scattering. It, however, will only influence the constant terms in the mode expansion and will not affect the impurity entropy. Therefore we conclude that the impurity entropy at the NR critical point is:
\begin{align}
g = 2 \sqrt{K}.
\end{align}
Here $2$ comes from the two possible occupations of the Majorana level. We see that at $K=1/4$ the entropy of the NR critical point becomes lower than the entropy of the AT critical point. This is consistent with the flow diagrams in Fig. \ref{fig:3d_phase_diagram}, where only for $K<1/4$ there is flow from the AT to the NR critical point.

\section{Derivation of the RG equations}

We derive the RG equations by first taking the harmonic part of the Hamiltonian, and then integrating out the high-energy modes. We thus rewrite the bosonic fields $\phi$ and $\theta$ at the normal transmission fixed point as follows:
\begin{align}
\phi = \phi_0(z, z^*) + \sum_{n=0}^N\frac{1}{\sqrt{2K\pi n}}&\left[e^{2\pi i n z/L}a_{n, R} \right. \\ &\left.+ e^{-2\pi i n z^*/L}a_{n, L} + h.c.\right],\nonumber\\
\theta = \theta_0(z, z^*) + \sum_{n=0}^N \sqrt{\frac{K}{2\pi n}}&\left[e^{2\pi i n z/L}a_{n, R} \right. \\ &\left.- e^{-2\pi i n z^*/L}a_{n, L} + h.c.\right].\nonumber
\end{align}
Here $N$ is the ultraviolet cutoff, which will be modified in RG, and $z=x+iu\tau$. $\phi_0$ and $\theta_0$ are abbreviations for zero modes, which will be irrelevant for the RG procedure. We rewrite this mode expansion via the integral over energies:
\begin{align}
\phi = \phi_0(z, z^*) + \int d\omega \sqrt{\frac{L}{K u \omega}}&\left[e^{i \omega z/u}a_{\omega, R} \right. \label{eq_supp:phi_expansion}\\ &\left.+ e^{-i \omega z^*/u}a_{\omega, L} + h.c.\right],\nonumber\\
\theta = \theta_0(z, z^*) + \int d\omega \sqrt{\frac{LK}{u \omega}}&\left[e^{i \omega z/u}a_{\omega, R} \right. \label{eq_supp:theta_expansion}\\ &\left.- e^{-i \omega z^*/u}a_{\omega, L} + h.c.\right].\nonumber
\end{align}
Indeed, if we plug the expression into the Luttinger liquid Hamiltonian \eqref{eq:H_bosonized}, we obtain the harmonic oscillator Hamiltonian for $a_{\omega, R/L}$:
\begin{align}
H_\omega = \frac{L}{u}\int \frac{d\omega}{2\pi}\omega(a_{\omega, R}^\dag a_{\omega, R} + a_{\omega, L}^\dag a_{\omega, L}).\label{eq_supp:harmonic}
\end{align}
Using this expression we can proceed with averaging the partition function $\exp\{-\beta H\}$, $\beta=1/k_B T$, over fast modes to obtain the low-energy Hamiltonian. For the second order in $t_1$ correction we find:
\begin{widetext}
\begin{align}
&\exp\{-\beta H_0\}\left\langle\mathcal{T} \left(\int d\tau t_1 d^\dag \Gamma (e^{i \sqrt{\pi} (\phi-(1- \tilde{U}_{NT}) \theta)} + e^{i \sqrt{\pi}(-\phi - (1-\tilde{U}_{NT})\theta)})(\tau) + h.c.\right )^2\right \rangle_>
\\
&=\exp\{- \beta H_0\}\left\langle \int d \tau_1 d\tau_2 \frac{t_1^2}{2} (d^\dag d - 1/2) \mathcal{T}\left[(e^{i \sqrt{\pi} (\phi-(1- \tilde{U}_{NT}) \theta)} + e^{i \sqrt{\pi}(-\phi - (1-\tilde{U}_{NT}) \theta)})(\tau_1) +  \right.\right. \\ &\times \left.\left.(e^{-i \sqrt{\pi}( \phi-(1- \tilde{U}_{NT}) \theta)}+ e^{-i \sqrt{\pi}(-\phi - (1-\tilde{U}_{NT}) \theta)})(\tau_2) + h.c. \right] + \Theta(\tau_{1}-\tau_2)\mathcal{T}\left[(e^{i \sqrt{\pi} (\phi-(1- \tilde{U}_{NT}) \theta)} + e^{i \sqrt{\pi}(-\phi - (1-\tilde{U}_{NT}) \theta)})(\tau_1) +  \right.\right. \\ &\times \left.\left.(e^{-i \sqrt{\pi}( \phi-(1- \tilde{U}_{NT}) \theta)}+ e^{-i \sqrt{\pi}(-\phi - (1-\tilde{U}_{NT}) \theta)})(\tau_2) - h.c. \right] \right\rangle_>\nonumber\\
&=\exp\{- \beta H_0\}\left\langle \int d \tau d\tau_{12} \frac{t_1^2}{2} (d^\dag d - 1/2) \mathcal{T} \left[(e^{i \sqrt{\pi} (\phi-(1- \tilde{U}_{NT}) \theta)} + e^{i \sqrt{\pi}(-\phi - (1-\tilde{U}_{NT}) \theta)})(\tau-\tau_{12}/2) \right.\right. \\ &\left.\left.(e^{-i \sqrt{\pi}( \phi-(1- \tilde{U}_{NT}) \theta)}+ e^{-i \sqrt{\pi}(-\phi - (1-\tilde{U}_{NT}) \theta)})(\tau+\tau_{12}/2) + h.c. \right]\right \rangle_>.\nonumber
\end{align}
Here the $\Theta$-function is due to time-ordering of the fermions, $\phi=\phi_> + \phi_<$, $\phi_>$ and $\theta_>$ are the fast modes we trace over to retain the Hamiltonian in terms of $\phi_<$, $\tau=(\tau_1+\tau_2)/2$, $\tau_{12} = (\tau_2 - \tau_1)$. In the last line we have dropped the $\Theta(\tau_{12})$ part as it enters with an anti-Hermitian operator and should integrate to zero (one can explicitly check that from the expressions below).  Since the d level has no dynamics in the unperturbed Hamiltonian and there is invariance under translation of $\tau$, we need to compute the averages:
\begin{align}
&\left\langle \int d\tau_{12} \mathcal{T} e^{i\sqrt{\pi}(\phi - (1 - \tilde{U}_{NT})\theta)}(0) e^{-i\sqrt{\pi}(\phi - (1 - \tilde{U}_{NT})\theta)}(\tau_{12})\right\rangle_>,\label{eq:average1}\\ &\left\langle \int d\tau_{12} \mathcal{T} e^{i\sqrt{\pi}(\phi - (1 - \tilde{U}_{NT})\theta)} (0) e^{i\sqrt{\pi}(\phi + (1 - \tilde{U}_{NT})\theta)}(\tau_{12})\right\rangle_>.\label{eq:average2}
\end{align}
The rest follows from these two by complex conjugation. We are interested in the connected parts of the averages to not double-count the first-order corrections to the coefficients. Thus we notice that these averages are converted to the difference of the connected and disconnected parts:
\begin{align}
\int d\tau_{12} \left(\exp\left\{-2 \pi \mathcal{G}_{s_1 s_2}(0, \tau_{12}) - 2 \pi \mathcal{G}_{s_1 s_2}(0, 0) \right\} - \exp \left\{-2\pi \mathcal{G}_{s_1 s_2}(0, 0)\right\}\right), \label{eq:average_appendix}
\end{align}
where $s_{1, 2}=\pm$, and
\begin{align}
\mathcal{G}_{s_1 s_2} (0, \tau_{12}) = \left\langle(\phi_> + s_1 (1-\tilde{U}_{NT})\theta_>)(0) (\phi_> + s_2 (1-\tilde{U}_{NT})\theta_>)(\tau_{12})\right\rangle.
\end{align}
In these expressions we hid the slow fields, since they are not averaged over. We will reintroduce them when we have computed the Green's functions above. For that we insert the well-known expressions for the Green's functions to obtain:
\begin{align}
\mathcal{G}_{s_1 s_2}(0, \tau_{12}) =& \int_{\Lambda'}^\Lambda \frac{d\omega}{\omega} e^{-\omega|\tau_{12}|} \left[\frac{1}{K}+ s_1 s_2 (1-\tilde{U}_{NT})^2 K + (s_1 + s_2) (1-\tilde{U}_{NT}) \Theta(\tau_{12})\right] \nonumber \\ \approx& \frac{d\Lambda}{\Lambda} e^{-\Lambda|\tau_{12}|} \left[\frac{1}{K}+ s_1 s_2 (1-\tilde{U}_{NT})^2 K + (s_1 + s_2) (1-\tilde{U}_{NT}) \Theta(\tau_{12})\right].
\end{align}
This leads to \eqref{eq:average_appendix} transforming to:
\begin{align}
\int d\tau_{12} \frac{-2\pi d\Lambda}{\Lambda}e^{-\Lambda|\tau_{12}|} \left[\frac{1}{K}+ s_1 s_2 (1-\tilde{U}_{NT})^2 K + (s_1 + s_2) (1-\tilde{U}_{NT})\Theta(\tau_{12})\right].
\end{align}

\end{widetext}

Finally, reintroducing the slow modes we notice that same signs in the exponents\eqref{eq:average1} and \eqref{eq:average2} are for the terms renormalizing the $U_{NT}$ term ($U_{NT} (d^\dag d - 1/2) \partial_x \phi(0)$), while opposite signs are producing the renormalization of the $U_{AR}$ term ($U_{AR} (d^\dag d - 1/2) \sin 2 \sqrt{\pi} \phi(0)$). Therefore $\delta U_{AR}$ is produced by the sum of $+-$ and $-+$ terms, while the $\delta U_{NT}$ is produced by the difference (negative sign comes from time-ordering) between $--$ and $++$ terms:
\begin{align}
\delta U_{AR} &= 4 \pi t_1^2 \left(\frac{1}{K} - K(1-\tilde{U}_{NT})^2\right) \frac{d\Lambda}{\Lambda^2}, \\
\delta U_{NT} &= 8 \pi t_1^2 (1-\tilde{U}_{NT}) \frac{d\Lambda}{\Lambda^2}.
\end{align}
At the same time we see that $U_{1,2}$ do not get renormalized.

Analogously we obtain renormalization of $t_1$ due to $U_{AR} t_1$ term of the form:
\begin{align}
\delta t_1 = 2 \pi t_1 U_{AR}\frac{1}{K} \frac{d\Lambda}{\Lambda^2}.
\end{align}

\section{Flow near the non-trivial fixed point}

In this section we confirm that the flow goes towards the non-trivial fixed point along two directions in the $(t,U_{NT}, U_{AR})$ space, and from it in one direction. We rewrite the RG equations using transformation $\tilde{t}^2 = \mathcal{X}$, $(1 - \tilde{U}_{NT})^2 = \mathcal{Y}$, and  $\tilde{U}_{AR}=\mathcal{Z}$. This transformation makes the RG equations second-order ones. We also use the $\epsilon$-expansion as above:
\begin{align}
\frac{d \mathcal{X}}{d \ell} &= 2 \mathcal{X} \left[1 - \frac{1}{4}\left(\frac{\mathcal{Y}}{ 4+4\epsilon}+4 + 4\epsilon\right)\right] + 2 \mathcal{X} \mathcal{Z}, \\
\frac{d \mathcal{Y}}{d\ell} &= - 8 \mathcal{Y} \mathcal{X},\\
\frac{d \mathcal{Z}}{d\ell} &=- (3 + 4\epsilon) \mathcal{Z} + \alpha \mathcal{X}.
\end{align}
Linearizing these equations near the fixed-point values $\mathcal{X}^0 = \epsilon(3 + 4\epsilon)/\alpha$, $\mathcal{Y}^0 = 0$, and $\mathcal{Z}^0 = \epsilon$, we obtain:
\begin{align}
\frac{d \delta\mathcal{X}}{d \ell} &= - \frac{\epsilon(3+4\epsilon)}{8\alpha(1+\epsilon)}\delta \mathcal{Y} + \frac{2 \epsilon(3+4\epsilon)}{\alpha} \delta\mathcal{Z}, \\
\frac{d \delta\mathcal{Y}}{d\ell} &= - \frac{8 \epsilon(3+4\epsilon)}{\alpha}\delta\mathcal{Y}, \\
\frac{d \delta\mathcal{Z}}{d\ell} &=- (3 + 4\epsilon) \delta\mathcal{Z} + \delta\mathcal{X}.
\end{align}
This system is solved by finding eigenvalues of the corresponding matrix:
\begin{align}
d\begin{pmatrix}
\delta \mathcal{X}\\
\delta \mathcal{Y}\\
\delta \mathcal{Z}
\end{pmatrix}/dt = \begin{pmatrix}
0 & - \frac{\epsilon(3+4\epsilon)}{8\alpha(1+\epsilon)} & \frac{\epsilon (6+8\epsilon)}{\alpha}\\
0& - \frac{8\epsilon(3+4\epsilon)}{\alpha} & 0 \\
1 & 0 & - (3+4\epsilon)
\end{pmatrix}
\begin{pmatrix}
\delta \mathcal{X}\\
\delta \mathcal{Y}\\
\delta \mathcal{Z}
\end{pmatrix}.
\end{align}
Straightforward calculations give the eigenvalues of the matrix in the first order in $\epsilon$: $-24 \epsilon/\alpha$, $-3 - (2+4\alpha)\epsilon/\alpha$, and $2\epsilon/\alpha$. The corresponding eigenvectors are: $\mathbf{a}_1=(3 - (24-4\alpha) \epsilon/\alpha, 208+(464-1536/\alpha)\epsilon, 1)^T$, $\mathbf{a}_2 = (-2\epsilon, 0, 1)^T$, and $\mathbf{a}_3 = (3+(2+4\alpha)\epsilon/\alpha, 0, 1)^T$. First two eigenvalues are negative, meaning flow to the critical point in the directions of $\mathbf{a}_1$ and $\mathbf{a}_2$, and the last one is positive, suggesting flow from the critical point in the direction of $\mathbf{a}_3$. This suggests that the separatrix plane, separating the NT and AR fixed points has a single critical point in it, to which the RG along the plane flows.

Finally, we obtain the equation for the separatrix by requiring that in the expansion of the vector $(\delta\mathcal{X}, \delta\mathcal{Y}, \delta\mathcal{Z})$ through $\mathbf{a}_i$ the coefficient in front of $\mathbf{a}_3$ vanishes. In the lowest order in $\epsilon$ this means:
\begin{align}
&\frac{\delta\mathcal{X}}{3} - \frac{\delta\mathcal{Y}}{208} + \frac{2\epsilon\delta \mathcal{Z}}{3\alpha} = 0, \\
&\frac{\tilde{t}^2 - 3\epsilon/\alpha}{3} - \frac{(1-\tilde{U}_{NT})^2}{208} + \frac{2\epsilon (\tilde{U}_{AR}-\epsilon)}{3\alpha} = 0.
\end{align}
We solve this to obtain the equation for the critical surface \eqref{eq:tildet}.

\section{Apparent discrepancy of the bosonic and fermionic RG}

When we look at the RG equations \eqref{eq:U(1)t}-\eqref{eq:AR}, we see that even when all the interactions are switched off, e.g. bare values of $U_{NT}$ and $U_{AR}$ are $0$, and $K=1$, the four-fermion couplings $U_{NT}$ and $U_{AR}$ get generated in RG procedure. At the same time if one starts with a fermionic model and integrate out high-energy degrees of freedom it is obvious that no four-fermion term can be generated and the Hamiltonian remains non-interacting. This leads to an apparent contradiction.

However, the contradiction is only apparent. There are several reasons for that. First, conceptually when one goes from the fermionic model to the bosonic model, though the transformation is exact, the high-energy modes in one model do not coincide with the high-energy modes of another model. Therefore the RG procedure is not obliged to give the exact same results for the operators present at intermediate stage. 

Second, the physical results of the RG procedure coincide with the solution of the non-interacting model. In our case this results in the same results for the stable fixed points for the bosonic and non-interacting fermionic solutions, besides the termperature exponent predicted by the bosonic calculation, $2$ for $K=1$, is the same as the one obtained in the non-interacting fermion solutions. Additional check comes from comparing the results of the interacting resonant level model (see \cite{Bou08} for example) with the solution of the non-interacting case. Again the bosonized version wields the same apparent problem as our RG, but the results for the critical points and exponents in the two cases coincide. 

Finally, we have checked that two- and four-fermion Green's functions computed in the bosonic and fermionic language after the renormalization coincide, thus proving that the contradiction is apparent but not present.

\end{document}